\documentclass[12pt,onecolumn,technote]{IEEEtran}

\usepackage{mathrsfs}
\usepackage{txfonts}
\usepackage [dvips] {graphicx}

\renewcommand\baselinestretch{2.0}

\ifCLASSINFOpdf

\else

\fi



\begin{document}

\title{Multi-Structural Signal Recovery for Biomedical Compressive Sensing}

\author{Yipeng~Liu*,~\IEEEmembership{Member,~IEEE},
        Maarten~De~Vos,~\IEEEmembership{Member,~IEEE},
        Ivan~Gligorijevic,
        Vladimir~Matic,
        Yuqian~Li,~\IEEEmembership{Student Member,~IEEE},
        and~Sabine~Van~Huffel,~\IEEEmembership{Fellow,~IEEE}% <-this % stops a space

\thanks{This work was supported by Research Council KUL: GOA MaNet, PFV/10/002 (OPTEC), IDO 08/013 Autism, several PhD/postdoc and fellow grants; Flemish Government: FWO: PhD/postdoc grants, projects: G.0427.10N (Integrated EEG-fMRI), G.0108.11 (Compressed Sensing) G.0869.12N (Tumor imaging) G.0A5513N (Deep brain stimulation), IWT: TBM070713-Accelero, TBM080658-MRI (EEG-fMRI), TBM110697-NeoGuard, PhD Grants, iMinds 2013, Flanders Care: Demonstratieproject Tele-Rehab III (2012-2014); Belgian Federal Science Policy Office: IUAP P719/ (DYSCO, `Dynamical systems, control and optimization', 2012-2017); ESA AO-PGPF-01, PRODEX (CardioControl) C4000103224; EU: RECAP 209G within INTERREG IVB NWE programme, EU HIP Trial FP7-HEALTH/ 2007-2013 (n¡ã 260777), EU MC ITN Transact 2012 $ \#  $ 316679; Alexander von Humboldt stipend. The scientific responsibility is assumed by its authors. \emph{Asterisk indicates corresponding author.}
}% <-this % stops a space
\thanks{Part of the paper was presented at the 34th Annual International Conference of the Engineering in Medicine and Biology Society (IEEE EMBC 2012), San Diego, USA, Aug. 28th - Sept. 1, 2012. }%
\thanks{*Y. Liu is with KU Leuven, Dept. of Electrical Engineering (ESAT), SCD-SISTA/iMinds Future Health Department, Kasteelpark Arenberg 10, box 2446, 3001 Heverlee, Belgium. (email: yipeng.liu@esat.kuleuven.be) }%
\thanks{M. De Vos is with Neuropsychology, Dept. of Psychology, University of Oldenburg, Oldenburg, Germany. (email: maarten.de.vos@uni-oldenburg.de) }
\thanks{I. Gligorijevic, V. Matic and S. Van Huffel are with KU Leuven, Dept. of Electrical Engineering (ESAT), SCD-SISTA/iMinds Future Health Department, Kasteelpark Arenberg 10, box 2446, 3001 Heverlee, Belgium. (email: ivan.gligorijevic@esat.kuleuven.be; vladimir.matic@esat.kuleuven.be; sabine.vanhuffel@esat.kuleuven.be) }%
\thanks{Y. Li is with the School of Electronic Engineering, University of Electronic Science and Technology of China (UESTC), Chengdu, 611731, China. (email: yuqianli@uestc.edu.cn)}
\thanks{Copyright (c) 2013 IEEE. Personal use of this material is permitted. However, permission to use this material for any other purposes must be obtained from the IEEE by sending an email to pubs-permissions@ieee.org.}}

\markboth{Journal Name,~Vol.~X, No.~X, Month~Year}%
{Shell \MakeLowercase{\textit{et al.}}: Bare Demo of IEEEtran.cls for Journals}

\maketitle

\begin{abstract}
%{\normalsize}

Compressive sensing has shown significant promise in biomedical fields. It reconstructs a signal
from sub-Nyquist random linear measurements. Classical methods only exploit the sparsity in one domain. A lot of biomedical signals have additional structures, such as multi-sparsity in different domains, piecewise smoothness, low rank, etc. We propose a framework to exploit all the available structure information. A new convex programming problem is generated with multiple convex structure-inducing constraints and the linear measurement fitting constraint. With additional \emph{a priori} information for solving the underdetermined system, the signal recovery performance can be improved. In numerical experiments, we compare the proposed method with classical methods. Both simulated data and real-life biomedical data are used. Results show that the newly proposed method achieves better reconstruction accuracy performance in term of both L1 and L2 errors.

\end{abstract}

\begin{IEEEkeywords}
%{\normalsize}

compressive sensing, biomedical signal reconstruction, sparsity, piecewise smoothness, low rank.

\end{IEEEkeywords}

\IEEEpeerreviewmaketitle

\section{Introduction}
\label{sec1}

\IEEEPARstart{C}{urrent} biomedical signals usually ask large amount of data to be sampled, transmitted, stored and processed. This results in large scale devices, time and power consumption \cite{Bachmann_wban} \cite{lustig_sparse_mri} \cite{lustig_cs_mri} \cite{guo_sparse_mrs}. Most of the current compression techniques sample the analog signal at the Nyquist rate, and then compress the data with different kinds of encoders. This acquisition process leads to a huge amount of irrelevant samples which are discarded during the compression stage of the signals. Besides, the high sampling rate requires a highly power-consuming analog-to-digital converter (ADC) with a large number of bits.

Compressive sensing (CS) can offer a solution. Rather than first sampling at a high rate and then compressing, it prefers to directly "sense" (acquire) the data in a compressive form at a much lower sampling rate \cite{candes_cs}. CS has attracted considerable attention in signal processing. It employs linear projections that preserve the structure of the signal as much as possible; the signal is then reconstructed from these projections using nonlinear signal recovery methods. It provides a new promising framework for acquiring signals.

Signal recovery is one of the key aspects of CS. Convex optimization is a popular way, due to its high recovery accuracy, guarantee of successful recovery, and the high availability of efficient algorithms. In the early stage of CS research, sparsity has been exploited by formulating an L1-norm based optimization problem. Sparsity is assumed in one domain as the key constraint to recover the signal \cite{candes_cs}. Recently, progress shows that other structure information can be exploited to recover signals \cite{chand_convex_geometry} \cite{chand_phd}, such as piecewise smoothness \cite{romberg_cs}, low-rank property \cite{fazel_rank_minimization}, \cite{candes_matrix_comletion}, orthogonality \cite{chand_phd}, permutation \cite{chand_phd}.

Much literature exists on CS applied to biomedical signal processing, such as magnetic resonance image (MRI), electromyography (EMG), electroencephalography (EEG), electrocardiography (ECG) \cite{lustig_sparse_mri} \cite{lustig_cs_mri} \cite{matic_sparse eeg}  \cite{mama_cs_ecg} \cite{addison_wavelet_ecg} \cite{salman_cs_emg} \cite{dixon_cs_emg}. However, most papers only exploit the sparsity in one signal domain, while many biomedical signals are sparse in more than one domain. Even more generally, some biomedical signals have structural features other than sparsity. For example, some EMG signals are sparse in both time and frequency domains \cite{salman_cs_emg} \cite{dixon_cs_emg}; Multi-channel EMG signals are highly-correlated with each other \cite{zwarts_multichannel-emg}, which can lead to a low-rank structure in the data matrix; MRI data have both a piecewise smooth structure and a low rank structure \cite{lustig_sparse_mri} \cite{candes_matrix_comletion}.

In this paper, we give a framework for exploiting multiple structures of biomedical signals for the recovery of the signal from sub-Nyquist samples with CS. First, we generalize the sparse signal model by allowing different kinds of possible structures applicable to biomedical signals. Then we incorporate all the available information about the data structures of the signal, by adding multiple convex structure-inducing constraints to enforce the corresponding structures in all the corresponding domains. By jointly constraining the multiple structure-inducing norms minimization and the data fitting, a new convex programming problem is presented for multi-structural signal recovery, which can be efficiently solved. As more \emph{a priori} information is used to solve the largely underdetermined system, the recovery performance is expected to be enhanced. Numerical experiments show the better performance of the proposed method compared to previous methods exploiting only one sparsity constraint, with both simulated and real-life biomedical data, such as block-sparse signals, ECG, EMG, MRI.

The major contributions of this paper can be summarized as below. First, sparsity was originally regarded as one of the two fundamental premises underlying CS. Classical methods only exploit the sparsity in one domain. Here we propose a novel signal recovery framework to exploit as many kinds of data structures as possible. In addition to the sparsity in one signal domain, other data structures are taken advantage too, such as sparsity in other domains, piecewise smoothness, low rank, etc. In CS, a small number of measurements are used to recover large scale data, which results in a largely underdetermined linear system. Hence, the signal recovery performance should be improved, provided the added regularizations are in accordance with the criterion used to judge the efficacy of a model. Second, we give a brief summary of the biomedical data structures and their representations. In the newly proposed framework, we propose three convex optimization models in CS applied to biomedical signals: L1-TV optimization for ECG signals, L1-L1 optimization for EMG signals, and L1-nuclear optimization for MRI. Generally, the used structure-inducing constraints, such as L1 norm minimization, total variation minimization, nuclear norm minimization, are based on previous investigations. They should be in accordance with the criterion used to judge the efficacy of a model. Numerical experiments also show that the proposed methods outperform the classical ones. Third, as far as we know, it is the first time that the cosparse signal recovery methods are used to recover biomedical signals from sub-sampled random measurements in CS. Besides their convenience to represent signals in the multi-structural signal recovery formulation, they have some other advantages, such as super-resolution, no incoherence requirement for the measurement matrix. Fourth, it is the first time that TV optimization is used to recover ECG signals. We show that the performance outperforms that of the classical sparse signal recovery methods.

The rest of the paper is organized as follows. Section \ref{sec2} presents the multi-structural signal model. In section \ref{sec3}, different kinds of norm regularizations for signal structures are discussed. In section \ref{sec4}, the convex programming problem for multi-structural signal recovery is presented. Numerical results are demonstrated in section \ref{sec6}. In section \ref{sec7}, we draw the conclusion.

\section{Multi-Structural Signal Model}
\label{sec2}

In a practical CS system, the analogue baseband signal \emph{x}(\emph{t}) is sampled using an analogue-to-information converter (AIC) \cite{laska_aic}. The AIC can be conceptually modeled as an ADC operating at Nyquist rate, followed by a sub-Nyquist linear operation. The random sub-Nyquist measurement vector $ {\bf{y}} \in {\textbf{R}^{M \times 1}} $ is obtained directly from the continuous-time signal \emph{x}(\emph{t}) by the AIC. For demonstration convenience, we formulate the sampling in discrete form as:

\begin{equation}
\label{eq2 measurement model_1}
{\bf{y}} = {\bf{\Phi x}}\
\end{equation}
where $ {\bf{\Phi }} \in \textbf{R} {^{M \times N}} $ is the measurement matrix (sensing matrix) with $ M \ll N $, and $ {\bf{x}} \in {\textbf{R}^{N \times 1}} $ is the sampled signal which can be regarded as the original signal obtained at Nyquist sampling rate.

Because in practice noise can not be avoided, the obtained sampling model with noise is:
\begin{equation}
\label{eq2 measurement model_2}
{\bf{y}} = {\bf{\Phi x}} + \textbf{n} \
\end{equation}
where \textbf{n} is the additive white Gaussian noise (AWGN) with zero mean and variance $ {\sigma ^2} $.

To enable CS, the measurement matrix $ {\bf{\Phi }} $ should satisfy some sufficient conditions, such as the restricted isometry property (RIP) \cite{baraniuk_rip}, the coherence condition \cite{tropp_omp}, the null space property (NSP) \cite{cohen_cs}, the constrained minimal singular values (CMSV) condition \cite{tang_cmsv}, etc. Usually one of the three types of measurement matrices are used: Gaussian matrix, Bernoulli matrix, or partial Fourier matrix.

The signal recovery from sub-Nyquist measurements is obviously an ill-posed inverse problem. The incorporation of prior information with a convex regulator is a popular way to deal with it. Such prior information specifies some simple signal structures. For biomedical signals, there are several common structures, such as sparsity, piecewise smoothness, low-rank property of the data matrix.

Sparsity exists in many biomedical signals. It means that many of the representation coefficients are close to or equal to zero, when the signal is represented in a certain domain. Traditionally, a representation model decomposes the signal into a linear combination of a few columns chosen from a predefined dictionary (representation matrix). Recently, a new signal model, called cosparse analysis model, was proposed \cite{nam_cosparsity}. In this new representation, an analysis operator multiplying the measurements leads to a sparse outcome. Let the signal in discrete form be expressed as:
\begin{equation}
\label{eq2 cosparse model}
{\bf{\theta }} = {\bf{\Psi x}}
\end{equation}
where $ {\bf{\Psi }} \in \textbf{R} {^{L \times N}} $ is the analysis operator (representation matrix/dictionary); $ {\bf{\theta }} \in {\textbf{R}^{L \times 1}} $ is the resulting sparse representative vector, i.e. most of the elements of $ {\bf{\theta }} $ are zero or almost zero. Here $ L \ge N $.

Besides sparsity, the processed signal has a piecewise smooth structure, in many biomedical signal processing applications \cite{romberg_cs}. The signal can be divided into several parts, and the adjacent elements of inner parts of every subsection are approximately smooth, while the elements on the boundaries of adjacent subsections can be quite different. For example, in MRI, an image often consists of several zones with abrupt boundaries between the zones.

Low rank is also a typical simple structural property of a signal matrix \cite{fazel_rank_minimization} \cite{candes_matrix_comletion} \cite{ivan_emg}, as originating from MRI, or multi-channel EMG. The rank of a matrix is its maximum number of linearly independent columns or rows. An $ L \times R $ matrix  $ {\bf{\Theta }} $ of rank \emph{K}, is called low-rank when $ K \ll \min (L,R) $. Its singular value decomposition (SVD) is given by:
\begin{equation}
\label{eq2 SVD}
{\bf{\Psi X}} = {\bf{\Theta }} = {\bf{U\Sigma }}{{\bf{V}}^H} = \sum\limits_{k = 1}^K {{\delta _k}{{\bf{u}}_k}{\bf{v}}_k^H}
\end{equation}
where $ {\bf{\Psi }} $ is an $ L \times N $ analysis operator, \textbf{X} is the $ N \times R $ signal matrix; $ {\bf{U}} = \left[ {\begin{array}{*{20}{c}}
{{{\bf{u}}_1}}&{{{\bf{u}}_2}}& \cdots &{{{\bf{u}}_L}}
\end{array}} \right] $  is an $ L \times L $ unitary matrix with $ {{\bf{u}}_l} $ being an $ L \times 1 $ column vector, the matrix $ {\bf{\Sigma }} $ is an $ L \times R $ diagonal matrix with nonnegative real numbers ${\delta _k}$, \emph{k} = 1, 2, ... , \emph{K} on the diagonal, and the $ R \times R $ unitary matrix ${{\bf{V}}^H}$  denotes the conjugate transpose of $ {\bf{V}} = \left[ {\begin{array}{*{20}{c}}
{{{\bf{v}}_1}}&{{{\bf{v}}_2}}& \cdots &{{{\bf{v}}_R}}
\end{array}} \right] $ with $ {{\bf{v}}_r} $ being an $ R \times 1 $ column vector. Recovering it from limited information is also a problem that has received considerable attention.

Considering the fact that some biomedical signals have more than one structural property simultaneously, their multi-structural model can be formulated as
\begin{equation}
\label{eq2 multi-sparse vector_model}
\begin{array}{*{20}{c}}
{{{\bf{\theta }}_1} = {{\bf{\Psi }}_1}{\bf{x}}}\\
{{{\bf{\theta }}_2} = {{\bf{\Psi }}_2}{\bf{x}}}\\
 \vdots \\
{{{\bf{\theta }}_P} = {{\bf{\Psi }}_P}{\bf{x}}}
\end{array}
\end{equation}
where \emph{P} is the number of analysis linear transformation matrices $ {{\bf{\Psi }}_p} $, \emph{p} = 1, 2, ... , \emph{P}. The corresponding expression in matrix form is:
\begin{equation}
\label{eq2 multi-sparse_matrix_model}
\begin{array}{*{20}{c}}
{{{\bf{\Theta }}_1} = {{\bf{\Psi }}_1}{\bf{X}}}\\
{{{\bf{\Theta }}_2} = {{\bf{\Psi }}_2}{\bf{X}}}\\
 \vdots \\
{{{\bf{\Theta }}_P} = {{\bf{\Psi }}_P}{\bf{X}}}
\end{array}
\end{equation}
By means of different linear transformation matrices, the resulting vectors $ {{\bf{\theta }}_p} $,  \emph{p} = 1, 2, ... , \emph{P} and matrices $ {{\bf{\Theta }}_p} $, \emph{p} = 1, 2, ... , \emph{P} have some simple and typical structural properties, such as sparsity, piecewise smoothness, low rank property, orthogonality.

\section{The Structure-Inducing Constraints}
\label{sec3}

After obtaining the random samples from AIC as in (\ref{eq2 measurement model_1}), the samples are processed in the digital signal processor (DSP) to recover the signal. Since $ M \ll N $, it is an ill-posed linear inverse problem. Since many biomedical signals have simple algebraic structures, such as the ones mentioned in section \ref{sec2}, some corresponding structure-inducing constraints can help to successfully recover the signals in combination with the linear measurement fitting error constraint. The problem can be formulated as:
\begin{equation}
\label{eq3 general_optimization}
\begin{array}{c}
\mathop {\min }\limits_{\bf{x}} f\left( {\bf{x}} \right)\\
{\rm{s}}{\rm{.~ t}}{\rm{.~~  }}\left\| {{\bf{y}} - {\bf{\Phi x}}} \right\| \le \varepsilon
\end{array}
\end{equation}
where $ f\left( {\bf{x}} \right) $ measures the degree of the structure of interest, and $ \varepsilon $  bounds the power of the AWGN in the measurements.

\subsection{Sparsity-inducing constraint}
\label{sec3.1}

A number of sparsity measures exist, such as L0 norm, L1 norm, normalized kurtosis, the Hoyer measure, Gini index, and so on \cite{hurley_sparsity_measure}. Minimization/maximization of one of them can encourage sparse structure in the recovered signal. The most commonly used and studied ones are the minimization of the L0 norm and L1 norm of the estimated signal. The L0 norm is defined as $ {\left\| {\bf{x}} \right\|_0} = \# \{ n:{x_n} \ne 0,n = 1,2, \cdots ,N\} $, which equals the number of nonzero elements of the vector $ {\bf{x}} = {\left[ {\begin{array}{*{20}{c}}
{{x_1}}&{{x_2}}& \cdots &{{x_N}}
\end{array}} \right]^T} $.

Using the L0 norm minimization to impose a sparse constraint in signal recovery yields
\begin{equation}
\label{eq3.1 L0_optimization}
\begin{array}{c}
\mathop {\min }\limits_{\bf{x}} {\left\| {{\bf{\Psi x}}} \right\|_0}\\
{\rm{s}}{\rm{.~t}}{\rm{.~~ }}{\bf{y}} = {\bf{\Phi x}}
\end{array}
\end{equation}
However, (\ref{eq3.1 L0_optimization}) is NP-hard unfortunately. One of the most popular ways to solve it is the basis pursuit (BP). It replaces the L0 norm with the L1 norm to yield a convex programming problem
\begin{equation}
\label{eq3.1 BP}
\begin{array}{c}
\mathop {\min }\limits_{\bf{x}} {\left\| {{\bf{\Psi x}}} \right\|_1}\\
{\rm{s}}{\rm{.~t}}{\rm{.~~ }}{\bf{y}} = {\bf{\Phi x}}
\end{array}
\end{equation}
where $ {\left\| {\bf{\theta }} \right\|_1} = \sum\nolimits_{n = 1}^N {\left| {{\theta _n}} \right|} $ is the L1 norm of the vector $ {\bf{\theta }} = {\left[ {\begin{array}{*{20}{c}}
{{\theta _1}}&{{\theta _2}}& \cdots &{{\theta _N}}
\end{array}} \right]^T} $. (\ref{eq3.1 BP}) can be solved efficiently by an interior-point method, subgradient algorithm \cite{boyd_cvx}, alternating direction method of multipliers (ADMM) \cite{boyd_admm}, and so on. Because it is a convex programming problem, it can guarantee efficient computation and global optimality.

To suppress the noise in measurements as shown in (\ref{eq2 measurement model_2}), the linear measurement fitting error constraint can be relaxed as done in the basis pursuit denoising (BPDN). It can be formulated as:
\begin{equation}
\label{eq3.1 BPDN}
\begin{array}{c}
\mathop {\min }\limits_{\bf{x}} {\left\| {{\bf{\Psi x}}} \right\|_1}\\
{\rm{s}}{\rm{.~ t}}{\rm{.~~  }}\left\| {{\bf{y}} - {\bf{\Phi x}}} \right\|_2^2 \le \varepsilon
\end{array}
\end{equation}

For block-sparse signals, L2/L1 optimization, which is a general case of the BPDN, is usually considered to recover the signal \cite{stojnic_block-sparse} \cite{eldar_block-sparse}. It can be formulated as:
\begin{equation}
\label{eq3.1 L2/L1 optimization}
\begin{array}{c}
\mathop {\min }\limits_{\bf{x}} \sum\limits_{d = 1}^D {\left\| {{{\bf{\Psi }}_d}{\bf{x}}} \right\|_2} \\
{\rm{s}}{\rm{.~ t}}{\rm{. ~~ }}\left\| {{\bf{y}} - {\bf{\Phi x}}} \right\|_2^2 \le \varepsilon
\end{array}
\end{equation}
where $ {\left\| {\bf{x}} \right\|_2} = \sqrt {\sum\nolimits_{n = 1}^N {{{\left| {{x_n}} \right|}^2}} } $ is the L2 norm of the vector
$ {\bf{x}} = {\left[ {\begin{array}{*{20}{c}}
{{x_1}}&{{x_2}}& \cdots &{{x_N}}
\end{array}} \right]^T} $; $ {{\bf{\Psi }}_d} $, \emph{d} = 1, 2, ... , \emph{D} is the \emph{d}-th block sub-dictionary, which gives birth to the \emph{d}-th block in the sparse signal, i.e.
\begin{equation}
\label{eq3.1 block-sparse_reprentation}
{{\bf{\theta }}_d} = {{\bf{\Psi }}_d}{\bf{x}};
\end{equation}

\begin{equation}
\label{eq3.1 block}
{{\bf{\theta }}^T} = \left[ {\begin{array}{*{20}{c}}
{{\bf{\theta }}_1^T}&{{\bf{\theta }}_2^T}& \cdots &{{\bf{\theta }}_D^T}
\end{array}} \right]
\end{equation}
where $ {{\bf{\theta }}_d} $, \emph{d} = 1, 2, ... , \emph{D} is the \emph{d}-th block of $ {\bf{\theta }} $. When \emph{D} = 1, L2/L1 optimization reduces to BPDN.

\subsection{Piecewise-smoothness-inducing constraint}
\label{sec3.2}
The piecewise smooth signal can have a sparse representation in the wavelet dictionary. However TV minimization is more popular to impose a piecewise smoothness constraint. Two TV formulations exist, i.e. TV1 and TV2 \cite{liu_tvs_beamformer} \cite{needell_tvm}:
\begin{equation}
\label{eq3.2 TV1}
{\left\| {\bf{x}} \right\|_{TV1}} = {\left\| {{\bf{Dx}}} \right\|_1}
\end{equation}

\begin{equation}
\label{eq3.2 TV2}
{\left\| {\bf{x}} \right\|_{TV2}} = {\left\| {{\bf{Dx}}} \right\|_2}
\end{equation}
where \textbf{D} is one of $ {{\bf{D}}_i},~i = 1,2, \cdots ,N $ as follows:

\begin{equation}
\label{eq3.2 differential matrix}
{{\bf{D}}_i} = \left[ {\begin{array}{*{20}{c}}
{{{\bf{D}}_{i,F}}}\\
{{{\bf{D}}_{i,B}}}
\end{array}} \right]
\end{equation}

\begin{equation}
\label{eq3.2 forward differential matrix}
{{\bf{D}}_{i,F}} = \left[ {\begin{array}{*{20}{c}}
{ - {\bf{1}}}&{\bf{1}}&0& \cdots &0&0&0\\
0&{ - {\bf{1}}}&{\bf{1}}& \cdots &0&0&0\\
 \vdots & \ddots & \ddots & \ddots & \ddots & \ddots & \vdots \\
0&0&0& \cdots &0&{ - {\bf{1}}}&{\bf{1}}\\
0&0&0& \cdots &0&0&{ - {\bf{1}}}
\end{array}} \right]
\end{equation}

\begin{equation}
\label{eq3.2 backward differential matrix}
{{\bf{D}}_{i,B}} = \left[ {\begin{array}{*{20}{c}}
{\bf{1}}&{ - {\bf{1}}}&0& \cdots &0&0&0\\
0&{\bf{1}}&{ - {\bf{1}}}& \cdots &0&0&0\\
 \vdots & \ddots & \ddots & \ddots & \ddots & \ddots & \vdots \\
0&0&0& \cdots &0&{\bf{1}}&{ - {\bf{1}}}\\
0&0&0& \cdots &0&0&{\bf{1}}
\end{array}} \right]
\end{equation}
$ {{\bf{D}}_{i,F}} $ and $ {{\bf{D}}_{i,B} }$ are the \emph{i}-th order forward and backward differential matrices; \textbf{1} is a $ 1 \times i $ row vector with all elements being one; and \textbf{-1}  is a $ 1 \times i $ row vector with all elements being -1. Usually the lengths of the vectors \textbf{1} and -\textbf{1} are set to 1. When used in TV1 (\ref{eq3.2 TV1}) and TV2 (\ref{eq3.2 TV2}), actually $ {{\bf{D}}_{i,F}} $ and $ {{\bf{D}}_{i,B} }$ would result into a similar expression. Hence, usually we only need to use one of them.

Generally TV1 is used more frequently than TV2. TV1 can be regarded as one kind of sparse constraint with the dictionary \textbf{D}.

Incorporation of the total variation minimization (TVM) constraints into the optimization model (\ref{eq3 general_optimization}) for signal recovery, yields

\begin{equation}
\label{eq3.2 TV optimization}
\begin{array}{c}
\mathop {\min }\limits_{\bf{x}} {\left\| {\bf{x}} \right\|_{TV}}\\
{\rm{s}}{\rm{.~ t}}{\rm{.~~  }}\left\| {{\bf{y}} - {\bf{\Phi x}}} \right\|_2^2 \le \varepsilon
\end{array}
\end{equation}
Here we call (\ref{eq3.2 TV optimization}) the TV optimization.

\subsection{Low-rank-inducing constraint}
\label{sec3.3}
To force a matrix to be of low rank, we can minimize the number of nonzero singular values. Inspired by the sparsity-inducing constraint, here we define:
\begin{equation}
\label{eq3.3 GS-p}
{\left\| {\bf{X}} \right\|_{GS - p}} = \left\{ {\begin{array}{*{20}{c}}
{{{\left( {\sum\limits_{i = 1}^{\min (M,N)} {\delta _i^p} } \right)}^{{1 \mathord{\left/
 {\vphantom {1 p}} \right.
 \kern-\nulldelimiterspace} p}}},~p \in \left( {0, + \infty } \right)}\\
{\# \{ {\delta _i} \ne 0,~i = 1,2, \cdots ,\min (M,N)\} ,~p = 0}
\end{array}} \right.
\end{equation}
where $ {\delta _i},~i = 1,2, \cdots ,\min (M,N) $ are the singular values of the matrix \textbf{X}. It is similar to the Schatten \emph{p}-norm which is defined as:
\begin{equation}
\label{eq3.3 S-p}
{\left\| {\bf{X}} \right\|_{S - p}} = {\left( {\sum\limits_{k = 1}^K {\delta _k^p} } \right)^{{1 \mathord{\left/
 {\vphantom {1 p}} \right.
 \kern-\nulldelimiterspace} p}}},p \in \left[ {1, + \infty } \right)
\end{equation}
We call (\ref{eq3.3 GS-p}) the generalized Schatten \emph{p}-norm, though it is not a real norm for $ 0 \le p < 1 $. $ {\left\| {\bf{X}} \right\|_{GS - 0}} $ is the best for measuring a low rank structure, but it is NP-hard. To improve efficiency when using a low-rank constraint, $ {\left\| {\bf{X}} \right\|_{GS - 0}} $ is relaxed to $ {\left\| {\bf{X}} \right\|_{GS - 1}} $ which is the well-known nuclear norm $ {\left\| {\bf{X}} \right\|_*} $ \cite{fazel_rank_minimization} \cite{candes_matrix_comletion}.

Combining the minimization of the nuclear norm with the data fitting error constraint, we write the problem as:
\begin{equation}
\label{eq3.3 convex low-rank recovery}
\begin{array}{c}
\mathop {\min }\limits_{\bf{X}} {\left\| {\bf{X}} \right\|_*}\\
{\rm{s}}{\rm{.~ t}}{\rm{.~~  }}{\bf{Y}} = {\bf{\Phi X}}
\end{array}
\end{equation}
(\ref{eq3.3 convex low-rank recovery}) is a convex programming problem, it can be solved efficiently. This nuclear norm based convex programming problem is often used for solving the matrix completion problem \cite{candes_matrix_comletion} \cite{hunyadi_nuc}.

Besides the constraints mentioned above, some other ones can be used, such as L2 norm minimization for Gaussian distribution structure, $ L\infty $ norm minimization for uniform distribution structure, spectral norm minimization for orthogonal matrix structure, and so on \cite{devos_combination}.

\section{Multi-Structural Signal Recovery}
\label{sec4}

To improve the recovery of compressively sampled biomedical signals, we can exploit the property that some biomedical signals have multiple structures simultaneously. For example, the ECG signal is piecewise smooth as can be seen in Fig. \ref{fig201} and sparse in the wavelet domain \cite{mama_cs_ecg} \cite{addison_wavelet_ecg}; EMG is sparse in both time domain and frequency domain \cite{salman_cs_emg} \cite{dixon_cs_emg}; MRI has a sparse representation and low-rank property \cite{romberg_cs} \cite{candes_matrix_comletion}, multi-channel EMG signals are sparse in some dictionaries and of low-rank \cite{zwarts_multichannel-emg} \cite{ivan_emg}. If properly used, the additional \emph{a priori} information can be helpful to improve the signal recovery performance.

Here we propose a new optimization model for multi-structural signal recovery as:
\begin{equation}
\label{eq4 multi-structure signal recovery}
\begin{array}{c}
\mathop {\min }\limits_{\bf{x}} \sum\limits_{p = 1}^P {{\lambda _p}{f_p}\left( {\bf{x}} \right)} \\
{\rm{s}}{\rm{.~ t}}{\rm{.~~  }}\left\| {{\bf{y}} - {\bf{\Phi x}}} \right\|_\diamondsuit^2 \le \varepsilon
\end{array}
\end{equation}
where \emph{P} is the number of analysis operators which generate structural outcomes; $ {\lambda _p} $ , \emph{p} = 1, 2, ... , \emph{P}, is the parameter balancing the different structure-inducing constraints, which can be tuned using cross validation \cite{hastie_statistical_learning}. $ {\left\| {\bf{x}} \right\|_\diamondsuit } $ is the L2 norm if \textbf{x} is a vector; and the Frobenius norm if \textbf{x} is a matrix. It is obvious that $ {\lambda _1}$  can be set to be 1. Here we call (\ref{eq4 multi-structure signal recovery}) multi-structure optimization. It is a scalarized formulation of multi-criterion optimization \cite{boyd_cvx}. Because all the used constraints $ {f_p}({\bf{x}}),{\rm{ }}p = 1,2, \cdots ,P $  are convex, efficient solutions exist, such as subgradient methods \cite{tropp_computational methods}, decomposition methods \cite{boyd_cvx}, ADMM \cite{boyd_admm}, and so on. Compared to the traditional ways which only use one kind of structural information, we give some examples when more \emph{a priori} information is used, and we expect to achieve better reconstruction performance.

We should note that the obtained optimal value is Pareto optimal. Since the multi-structural signal recovery is a multi-criterion optimization problem, we know that the optimal values could not be the same for all criteria. In practice, the choice of the $ \lambda _p $ may be dependent. Similarly to the choice of the parameter $ \varepsilon $ in BPDN (\ref{eq3.1 BPDN}), the optimal choice is dependent on the true solution \textbf{x}, and is therefore difficult to obtain. For this reason, various sub-optimal approaches exist for the selection. One way can be to select a fixed one based on experience. The other way is learning. In numerical experiments, the training data can be generated. Cross validation is a simple and widely used learning way. Instead of using the entire data set when training a learner, some of the data is removed prior to training. After training, the removed data can be used to test the performance of the learned model on "new" data \cite{hastie_statistical_learning}.

\subsection{L1-TV optimization}
\label{sec4.1}
For piecewise smooth and sparse signals, we combine the L1 norm minimization constraint and TVM constraint in the multi-structure optimization problem (\ref{eq4 multi-structure signal recovery}). We set \emph{P} = 2, $ {f_1}\left( {\bf{x}} \right) = {\left\| {\bf{x}} \right\|_{TV}} $ and $ {f_2}\left( {\bf{x}} \right) = {\left\| {{\bf{\Psi x}}} \right\|_1} $. The multi-structure optimization problem (\ref{eq4 multi-structure signal recovery}) reduces to:
\begin{equation}
\label{eq4.1 L1-TV optimization}
\begin{array}{c}
\mathop {\min }\limits_{\bf{x}} \left( {{{\left\| {\bf{x}} \right\|}_{TV}} + {\lambda _2}{{\left\| {{\bf{\Psi x}}} \right\|}_1}} \right)\\
{\rm{s}}{\rm{.~ t}}{\rm{.~~  }}\left\| {{\bf{y}} - {\bf{\Phi x}}} \right\|_2^2 \le \varepsilon
\end{array}
\end{equation}
We call (\ref{eq4.1 L1-TV optimization}) the L1-TV optimization. Here we generalize the TV constraint by taking a linear combination of TV1 and TV2 constraints as $ {\left\| {\bf{x}} \right\|_{TV1}} + \lambda {\left\| {\bf{x}} \right\|_{TV2}} $, where $ \lambda $ is a scalar balancing the two constraints.

In the special case of block-sparse signals, another form of multi-structure signal recovery can be formulated by replacing the L1 norm  with L2/L1 mixed norm. The obtained optimization is:
\begin{equation}
\label{eq4.1 L2/L1-TV optimization}
\begin{array}{c}
\mathop {\min }\limits_{\bf{x}} \left( {{{\left\| {\bf{x}} \right\|}_{TV}} + {\lambda _2}\sum\limits_{d = 1}^D {\left\| {{{\bf{\Psi }}_d}{\bf{x}}} \right\|_2} } \right)\\
{\rm{s}}{\rm{.~ t}}{\rm{.~~  }}\left\| {{\bf{y}} - {\bf{\Phi x}}} \right\|_2^2 \le \varepsilon
\end{array}
\end{equation}
We call (\ref{eq4.1 L2/L1-TV optimization}) the L2/L1-TV optimization.

\subsection{L1-L1 optimization}
\label{sec4.2}

The multi-structure optimization can also be applied to the signals that are nearly sparse in multiple domains. For example, to reconstruct the EMG signals which are sparse in both time and frequency domains, we set \emph{P} = 2, $ {f_1}\left( {\bf{x}} \right) = {\left\| {\bf{x}} \right\|_1} $ and $ {f_2}\left( {\bf{x}} \right) = {\left\| {{\bf{Fx}}} \right\|_1} $. \textbf{F} is the discrete Fourier transformation (DFT) matrix. Generally a signal can not be sparse in both time and frequency domains. But the sparsity here does not strictly refer to the number of nonzero elements but to the number of significantly small elements. The multi-structure optimization for EMG signal recovery can be reformulated as \cite{liu_multi-sparsity}
\begin{equation}
\label{eq4.2 L1-L1 optimization}
\begin{array}{c}
\mathop {\min }\limits_{\bf{x}} \left( {{{\left\| {\bf{x}} \right\|}_1} + {\lambda _2}{{\left\| {{\bf{Fx}}} \right\|}_1}} \right)\\
{\rm{s}}{\rm{.~ t}}{\rm{. ~~ }}\left\| {{\bf{y}} - {\bf{\Phi x}}} \right\|_2^2 \le \varepsilon
\end{array}
\end{equation}
(\ref{eq4.2 L1-L1 optimization}) is called L1-L1 optimization.

\subsection{L1-nuclear optimization}
\label{sec4.3}
Many biomedical images are sparse in some domains and of low rank simultaneously, such as different kinds of MRI. It also applies to many multi-channel biomedical signals with highly correlated channels, such as multi-channel EMG, etc. To recover this kind of signal, we set \emph{P} = 2, $ {f_1}\left( {\bf{X}} \right) = {\left\| {{\mathop{\rm vec}\nolimits} ({\bf{\Psi X}})} \right\|_1} $ and $ {f_2}\left( {\bf{X}} \right) = {\left\| {\bf{X}} \right\|_*} $, and we can get:

\begin{equation}
\label{eq4.3 L1-nuclear optimization}
\begin{array}{c}
\mathop {\min }\limits_{\bf{X}} \left( {{{\left\| {{\mathop{\rm vec}\nolimits} ({\bf{\Psi X}})} \right\|}_1} + {\lambda _2}{{\left\| {\bf{X}} \right\|}_*}} \right)\\
{\rm{s}}{\rm{.~ t}}{\rm{. ~~ }}\left\| {{\bf{Y}} - {\bf{\Phi X}}} \right\|_F^2 \le \varepsilon
\end{array}
\end{equation}
where $ {\mathop{\rm vec}\nolimits} ({\bf{X}}) $  puts all the columns of \textbf{X} into one vector; and $ {\left\| {\bf{X}} \right\|_F} $ is the Frobenius norm of the matrix \textbf{X}. (\ref{eq4.3 L1-nuclear optimization}) is called L1-nuclear optimization.

\vspace{1\baselineskip}

The general formulation for convex optimization can be written as
\begin{equation}
\label{eq4 convex_optimization}
\begin{array}{l}
\min~ {\rm{   }}{f_0}({\bf{x}}),{\rm{ }}\\
{\rm{s}}{\rm{.~ t}}{\rm{.~~   }}{f_i}({\bf{x}}) \le {b_i},{\rm{ }}i = 1, \cdots ,M
\end{array}
\end{equation}
where the variable \textbf{x} is of length \emph{N}. The computational time is roughly proportional to $ \max \{ {N^3},{N^2}M,G\} $, where \emph{G} is the cost for evaluating the functions $ {f_i} $ and their first and second derivatives \cite{boyd_cvx}. Additional regularizers ask for more computations, which makes \emph{G} larger. But compared with the single structure-inducing constraint based optimization problem, the additional computational complexity should not be significant. Considering the accuracy performance improvement, it should be worthwhile. For example, if we used the subgradient methods to solve the convex optimization problem, one subgradient of an L1 norm $ {\left\| {{\bf{Ax}}} \right\|_1} $ is $ {{\bf{A}}^T}{\mathop{\rm sgn}} \left( {{\bf{Ax}}} \right) $, where $ {\bf{A}} \in {\textbf{R}^{M \times N}} $  and $ {\bf{x}} \in {\textbf{R}^{N \times 1}} $ \cite{boyd_cvx}. The additional computation time should be approximiatedly proportional to $ 4MN $ for each iteration step. Compared with \emph{M} and \emph{N}, the number of iteration steps to convergence should be smaller, because in the length of the signal \emph{N} should be considerablely large where CS is used. Therefore, compared with $ \max \{ {N^3},{N^2}M,G\} $, the additional computation time is not very large and can be acceptable.

\section{Numerical Experiments}
\label{sec6}

To quantify the performance of signal recovery, the estimation errors are calculated via the following formulas:
\begin{equation}
\label{eq6_vector error}
e = \frac{1}{C}\sum\limits_{c = 1}^C {{{\left\| {{{\bf{x}}_c} - {{{\bf{\hat x}}}_c}} \right\|}_b}}
\end{equation}
for vectors $ {{\bf{x}}_c} $ and $ {{\bf{\hat x}}_c} $; and

\begin{equation}
\label{eq6_matrix error}
e = \frac{1}{C}\sum\limits_{c = 1}^C {{{\left\| {{\bf{vec}}\left( {{{\bf{X}}_c} - {{{\bf{\hat X}}}_c}} \right)} \right\|}_b}}
\end{equation}
for matrices $ {{\bf{X}}_c} $ and $ {{\bf{\hat X}}_c} $, where $ {{\bf{x}}_c} $ and $ {{\bf{X}}_c} $ are the original signals in the \emph{c}-th simulation; $ {{\bf{\hat x}}_c} $ and $ {{\bf{\hat X}}_c} $ are the estimated signals in the \emph{c}-th simulation; \emph{C} is the number of simulations.  $ b \in \{ 1,2\}  $ indicates the criteria. When \emph{b} = 1, it represents the mean L1 error; and when \emph{b} = 2, it represents the mean L2 error.

To demonstrate the performance improvement of the proposed multi-structural optimization for biomedical signals, we perform four groups of numerical experiments. The first group uses L1-TV optimization and L2/L1-TV1 optimization to recover some simulated signals; the second group uses L1-TV1-TV2 optimization to reconstruct the ECG signals; the third group uses L1-L1 optimization to recover the EMG signals; in the third group, MRIs are reconstructed by L1-nuclear optimization. In each group of experiments, some related methods are used for comparison, such as least squares (LS) methods, BPDN, nuclear norm based matrix recovery.

In numerical experiments, the K-fold cross validation is used to learn the parameters $ {\lambda _p} $, \emph{p} =1, 2, ... , \emph{P} \cite{hastie_statistical_learning}. The training data can be generated, because the original data are available, and the compressed measurements can be obtained by the product of the measurement matrix and the original data as (\ref{eq2 measurement model_1}). We can generate \emph{T} groups of data. \emph{T} iterations of training and validation are performed. In each iteration, we only use \emph{T}-1 groups of data (training subset) for training, and use the remaining group of data for validation. In the training of each interation, we choose the optimal $ {\lambda _{p,t}} $, \emph{p} =1, 2, ... , \emph{P}, \emph{t} = 1, 2, ... , \emph{T}-1 to achieve the smallest residual $ {r_t} = {\left\| {{{{\bf{\hat x}}}_t} - {{\bf{x}}_t}} \right\|_2} $ by exhaustive searching method. Then we get the average $ {\bar \lambda _p} = \frac{1}{{T - 1}}\sum\nolimits_{t = 1}^{T - 1} {{\lambda _{p,t}}} $ , and $ {r_{training}} = \frac{1}{{T - 1}}\sum\nolimits_{t = 1}^{T - 1} {{r_t}} $. With the learned $ {\bar \lambda _p} $ , \emph{p} =1, 2, ... , \emph{P}, we can use the remaining group of data to test whether the testing residual approximately equals the average training residual. i. e. $ \left| {{r_{testing}} - {r_{training}}} \right| \le \delta \left| {{r_{training}}} \right| $, where $ \delta  \ge 0 $ is a small scalar. The 10-fold cross validation (\emph{T} = 10) is the most common, and is used here.

EEG signals are another typical class of biomedical signals. However, to our knowledge, they don't have any data structure except sparsity. In fact, even the sparsity of EEG is controversial. \cite{zhang_EEG_not_sparse} claims that EEG is non-sparse in the time domain and also non-sparse in transformed domains (such as the wavelet domain). This was also verified in our experiments based on our EEG data. Therefore, currently we can not use the proposed method to recover an EEG signal from its compressive measurements.

\subsection{Simulated signals}
\label{sec6.1}
\begin{figure}[!h]
 \centering
 \includegraphics[angle= 0, scale = 0.32]{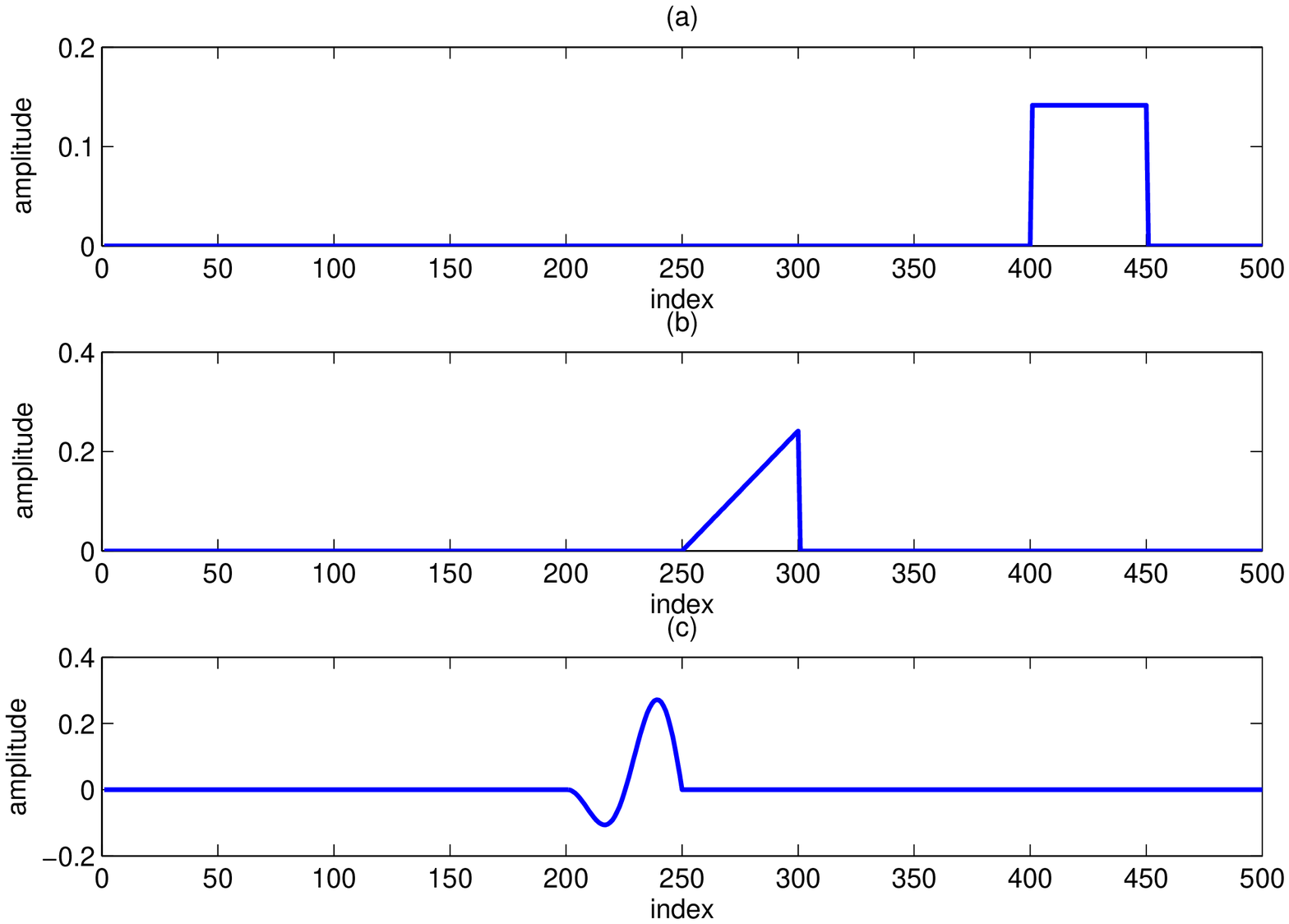}
 \caption{Three types of block sparse signals with different inner block structures.}
 \label{fig101}
\end{figure}

In the first group of numerical experiments, we simulated a series of signals which are sparse and piecewise smooth simultaneously. The length of the signal is \emph{N} = 500. The number of measurements ranges from \emph{M} = 10 to \emph{M} = 100. The measurement matrix consists of the entries sampled from an i.i.d. Gaussian distribution. The signal-to-noise ratio (SNR) of the signal is 5. Every signal is normalized by its L2 norm. Fig. \ref{fig101} gives several examples of the signals. As shown, there is a nonzero block, randomly positioned in the signal. The width of the block is 50. Inside the block, the elements can be constant, linearly increasing, or sinusoidal. The number of Monte Carlo simulations is set to be 1000, i.e. \emph{C} = 1000. Five methods are employed to recover the signals. They are BP, L2/L1 optimization, TV optimization, L1-TV optimization and L2/L1-TV optimization. The TV herein refers to TV1.

\begin{figure}[!h]
 \centering
 \includegraphics[scale = 0.32]{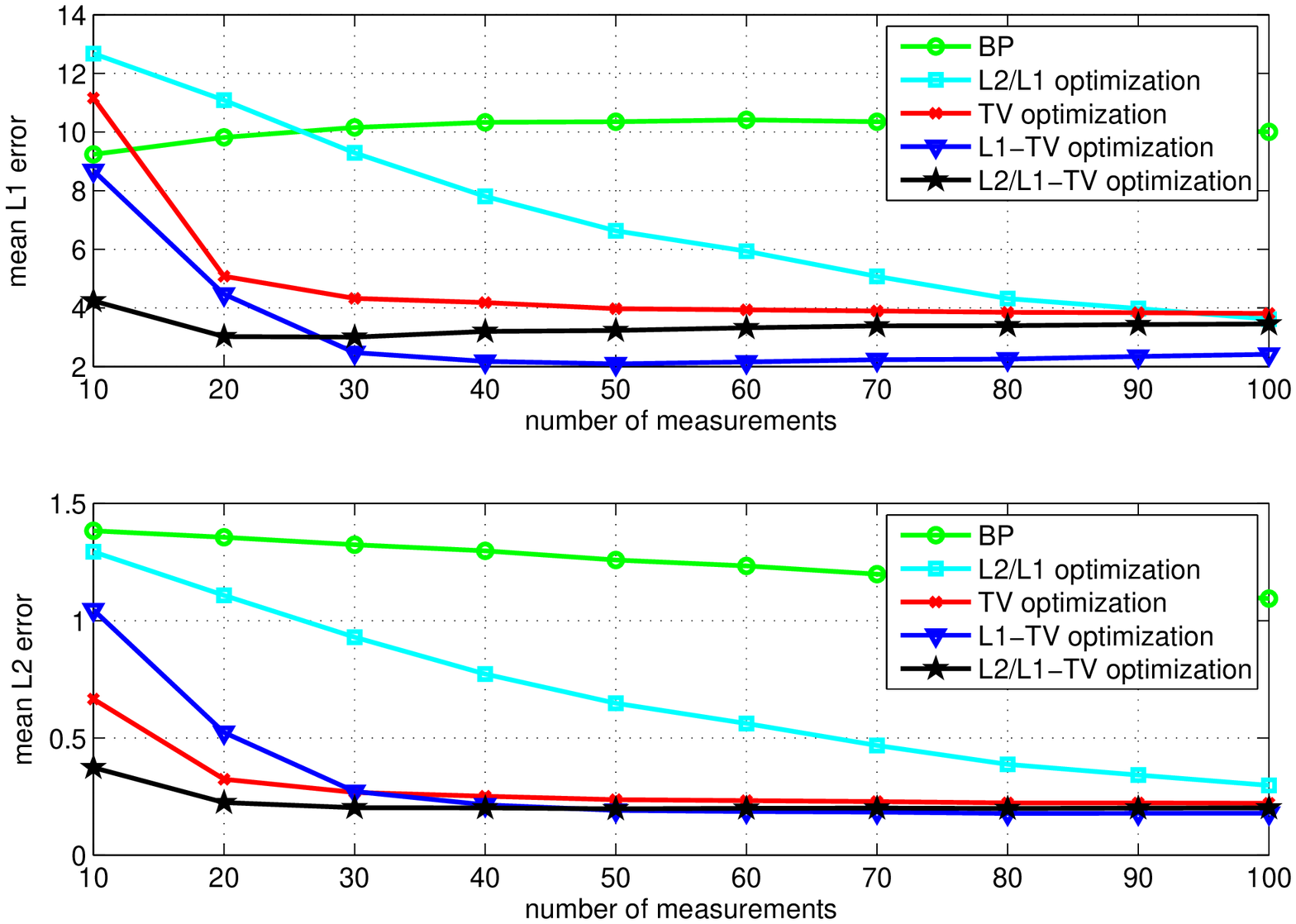}
 \caption{The mean L1 and L2 errors versus the number of measurements with different kinds of recovery methods when the block is a rectangle as in Fig. \ref{fig101}a.}
 \label{fig102}
\end{figure}

\begin{figure}[!h]
 \centering
 \includegraphics[scale = 0.32]{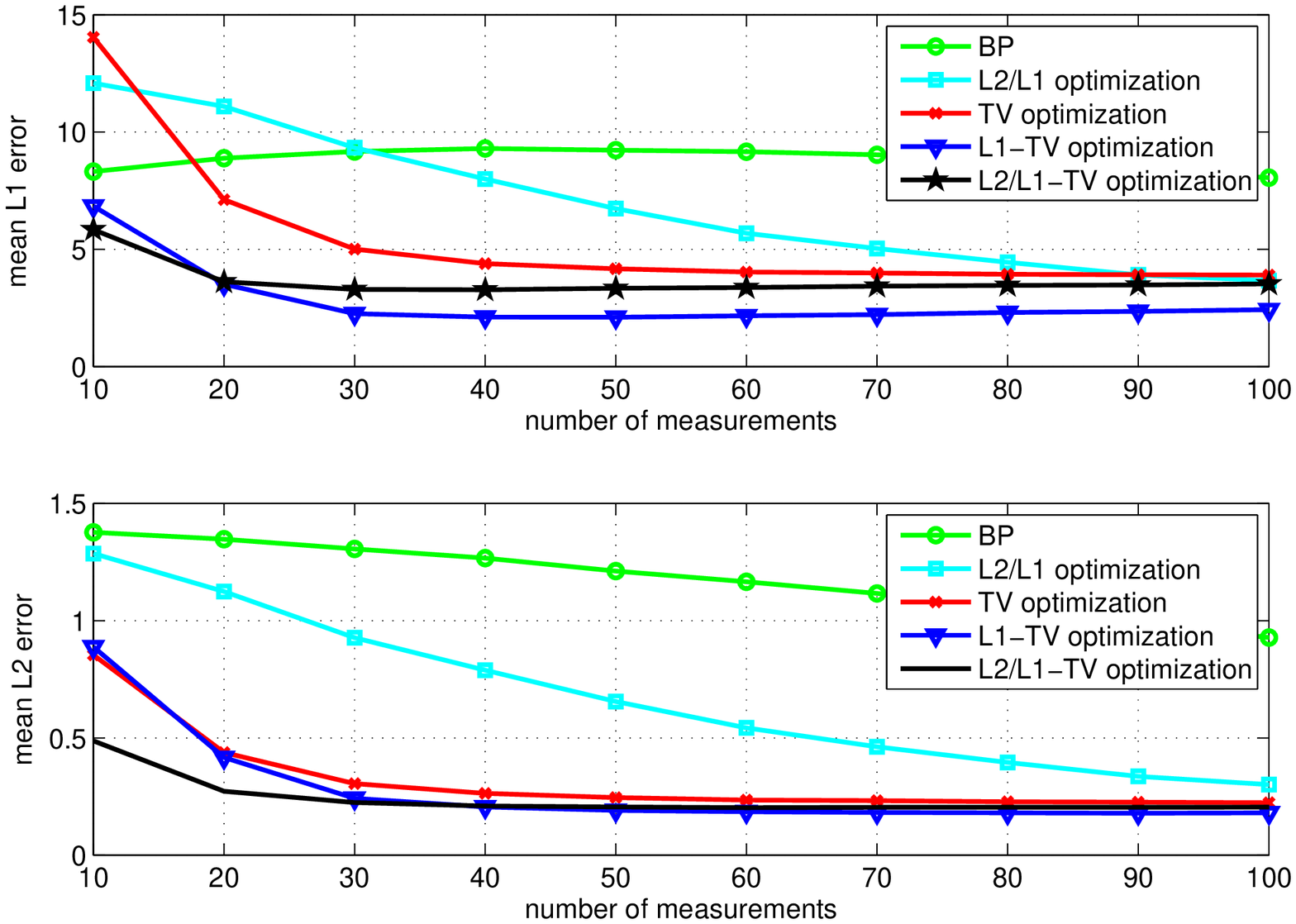}
 \caption{The mean L1 and L2 errors versus the number of measurements with different kinds of recovery methods when the block is a triangle as in Fig. \ref{fig101}b.}
 \label{fig103}
\end{figure}

\begin{figure}[!h]
 \centering
 \includegraphics[scale = 0.32]{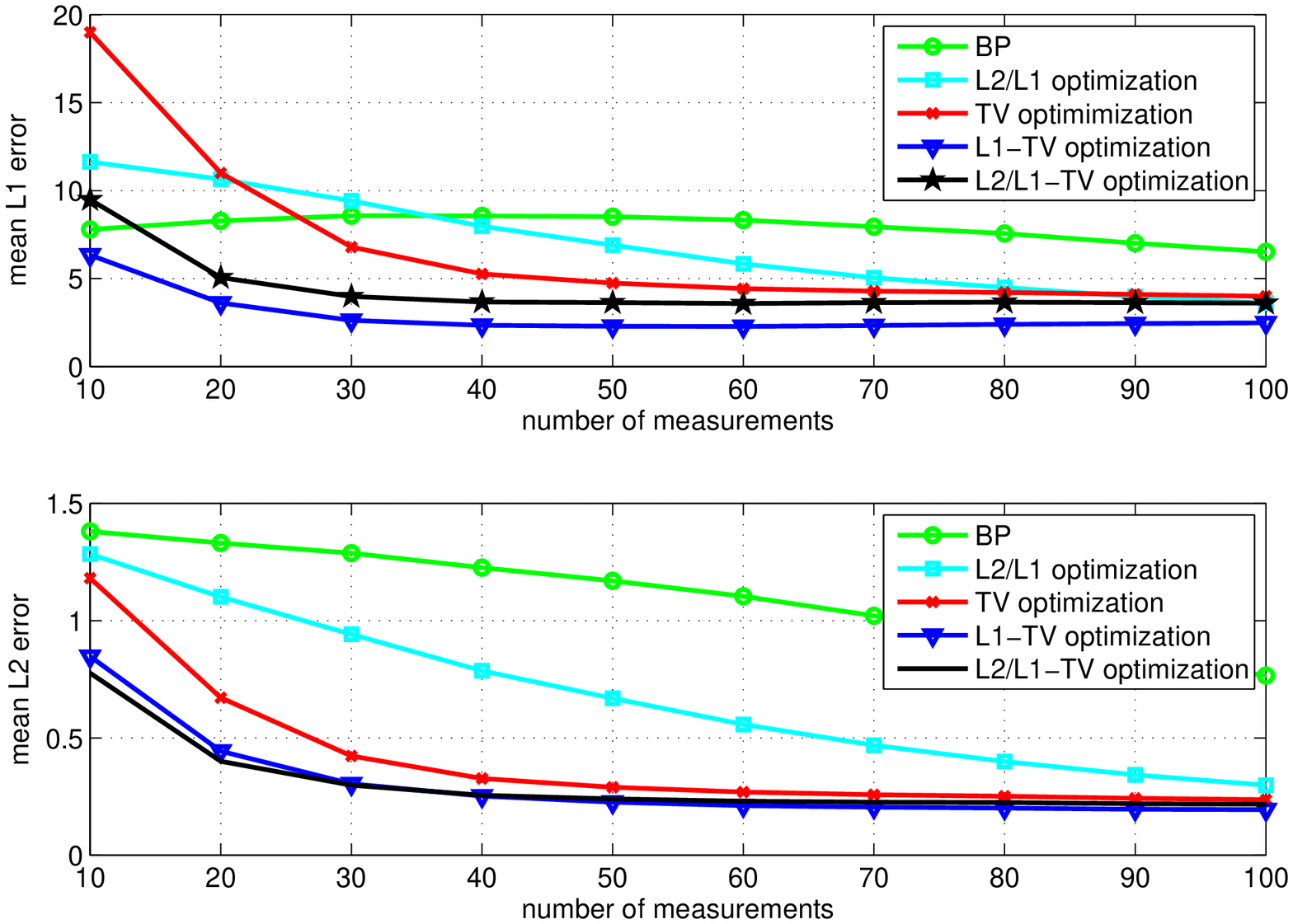}
 \caption{The mean L1 and L2 errors versus the number of measurements with different kinds of recovery methods when the block is a period of sine waveform as in Fig. \ref{fig101}c.}
 \label{fig104}
\end{figure}

\begin{figure}[!h]
 \centering
 \includegraphics[angle= 0, scale = 0.28]{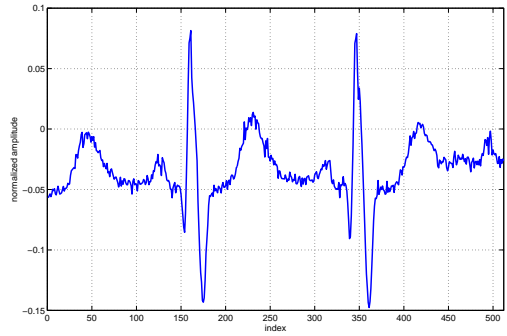}
 \caption{One section of the used ECG signal.}
 \label{fig201}
\end{figure}

Fig. \ref{fig102} gives the mean L1 and L2 errors for simulated constant block signals in 1000 Monte Carlo simulations; Fig. \ref{fig103} gives the mean L1 and L2 errors for simulated triangle block signals in 1000 Monte Carlo simulations; and Fig. \ref{fig104} gives the mean L1 and L2 errors for simulated sine block signals in 1000 Monte Carlo simulations. It is obvious that the two multi-structural optimization methods, L1-TV optimization and L2/L1-TV optimization, outperform the others. The two multi-structural optimization methods almost achieve the same mean L2 error performance. The L1-TV optimization is better with mean L1 error performance.

\subsection{ECG signals}
\label{sec6.2}

The used ECG data is obtained from the \emph{Physiobank} database \cite{glodberger_physiobank} \cite{bousseljot_ecg}. 9600 measurements are uniformly obtained in one hour and used as the original signal.
Previously used signal recovery methods are mainly based on sparse signal recovery methods, such as BP, and orthogonal matching pursuit (OMP) \cite{mama_cs_ecg}. Here we propose to exploit ECG signal's piecewise smoothness property by using the TV optimization. We use the L1-TV optimization to make use of both piecewise smoothness and sparsity in the wavelet domain. The L1-TV optimization for ECG signals is compared with other single structure constraint methods, such as BP, TV optimization. The utilized dictionary is given by the
orthogonal Daubechies wavelets (db 10) which is reported to be the most popular wavelet family for ECG compression \cite{mama_cs_ecg}. Here we divide the obtained ECG signal into sections. The length of every section is \emph{N} = 512. Fig. \ref{fig201} shows one section of the ECG signals. The number of measurements used ranges from \emph{M} = 20 to \emph{M} = 300. The elements of the measurement matrix are i.i.d. sampled from a Gaussian distribution. Every section of the signal is normalized by its L2 norm.

\begin{figure}[!h]
 \centering
 \includegraphics[scale = 0.32]{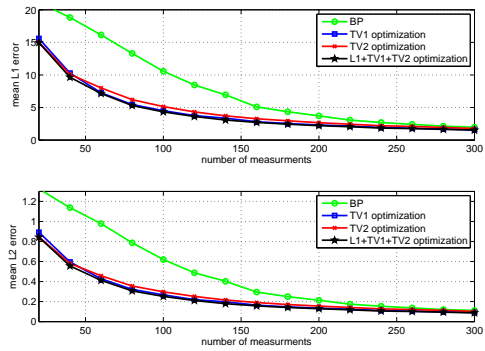}
 \caption{The mean L1 and L2 errors versus the number of measurements with different kinds of methods for ECG signal recovery.}
 \label{fig202}
\end{figure}

\begin{figure}[!h]
 \centering
 \includegraphics[scale = 0.32]{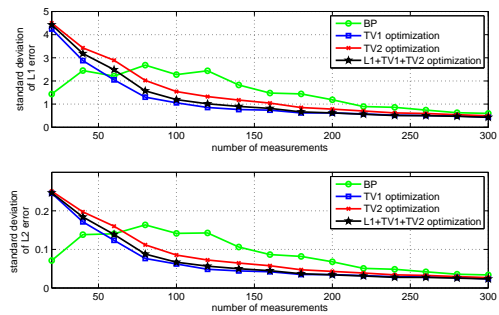}
 \caption{The standard deviation of the L1 and L2 errors versus the number of measurements with different kinds of methods for ECG signal recovery.}
 \label{fig203}
\end{figure}

Fig. \ref{fig202} gives the mean L1 and L2 errors and Fig. \ref{fig203} gives the standard deviation of L1 and L2 errors with the number of measurements ranging from \emph{M} = 20 to \emph{M} = 300 in \emph{C} = 280 simulations; We can see that the performance of BP is far worse than the others. Comparing the other three methods, the L1-TV1-TV2 optimization achieves the smallest mean L1 and L2 errors from \emph{M} = 20 to \emph{M} = 300.  It gives the best performance indeed, though the improvement is not significant. The L1-TV1-TV2 optimization and the TV1 optimization have nearly the same standard deviation performance when the number of measurements is larger than 100, i.e. when the mean errors are considerably acceptable.

\subsection{EMG signals}
\label{sec6.3}

\begin{figure}[!h]
 \centering
 \includegraphics[scale = 0.32]{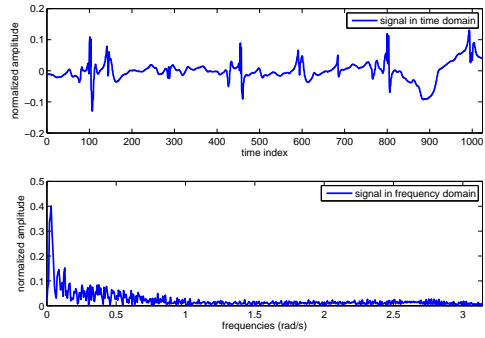}
 \caption{An example of EMG data from a healthy person: $ EMG - healthy $.}
 \label{fig301}
\end{figure}

The EMG signals are obtained from the \emph{Physiobank} database \cite{glodberger_physiobank} too. Data were collected with a Medelec Synergy N2 EMG Monitoring System (Oxford Instruments Medical, Old Woking, United Kingdom). A 25mm concentric needle electrode was placed into the tibialis anterior muscle of each subject. The patient was then asked to dorsiflex the foot gently against resistance. The needle electrode was repositioned until motor unit potentials with a rapid rise time were identified. Data were then collected for several seconds, after which the patient was asked to relax and the needle removed. Fig. \ref{fig301} shows three examples of EMG data from: a) a 44 year old man without history of neuromuscular disease; b) a 62 year old man with chronic low back pain and neuropathy due to a right L5 radiculopathy; and c) a 57 year old man with myopathy due to longstanding history of polymyositis, treated effectively with steroids and low-dose methotrexate. The data were recorded at 50 KHz and then downsampled to 4 KHz. During the recording process two analog filters were used: a 20 Hz high-pass filter and a 5K Hz low-pass filter.

In \cite{salman_cs_emg}, the static thresholding algorithm is used to reconstruct the EMG signals. But those thresholding methods are proved to be worse than convex relaxation. The measurement matrix  $ {\bf{\Phi }} $ is formed by sampling the i.i.d. entries from a white Gaussian distribution. Here four signal recovery methods, namely the Least Squares (LS) methods with $ {\min _{\bf{x}}}{\left\| {\bf{x}} \right\|_2},~{\rm{ s}}{\rm{.~t}}{\rm{.~}}\left\| {{\bf{y}} - {\bf{\Phi x}}} \right\| \le \varepsilon $, BPDN with dictionary the identity matrix (T-L1 optimization), BPDN with dictionary the DFT matrix (F-L1 optimization), and the newly proposed L1-L1 optimization with both the identity matrix and DFT matrix as the dictionaries,  are used to reconstruct the EMG signals. Both T-L1 optimization and F-L1 optimization are in the form of BPDN.  $ {\lambda _2} $ is chosen to be 0.05; $ \varepsilon $ is chosen to be $ 5\% $ of the measurement power, i. e. $ \varepsilon  = 0.05{\left\| {\bf{y}} \right\|_2} $. Because the amount of available data is limited, the number of simulations \emph{C} is chosen to be 40 here. Every section of the signal is normalized by its L2 norm.

\begin{figure}[!h]
 \centering
 \includegraphics[scale = 0.32]{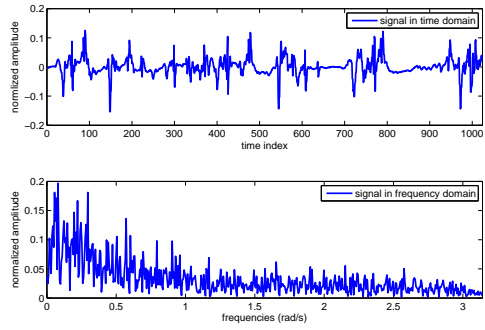}
 \caption{An example of EMG data from a patient with myopathy: $ EMG - myopathy $.}
 \label{fig302}
\end{figure}

\begin{figure}[!h]
 \centering
 \includegraphics[scale = 0.32]{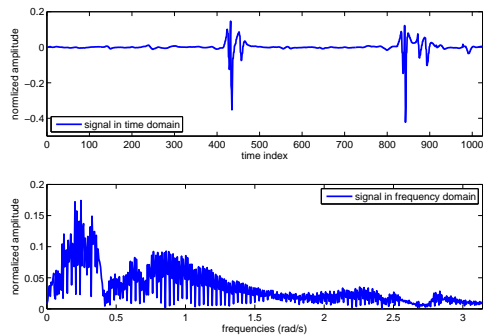}
 \caption{An example of EMG data from a patient with neuropathy: $ EMG - neuropathy $.}
 \label{fig303}
\end{figure}

Fig. \ref{fig301}, Fig. \ref{fig302} and Fig. \ref{fig303} show three sections of EMG signals of a healthy person ($ EMG - healthy $), a patient with myopathy ($ EMG - myopathy $) and a patient with neuropathy ($ EMG - neuropathy $), respectively. We can see that all three signals are sparse in the time domain. In the frequency domain, $ EMG - healthy $ and $  EMG - myopathy $ signals are sparse but the $ EMG - neuropathy $ signal is not.

\begin{figure}[!h]
 \centering
 \includegraphics[scale = 0.32]{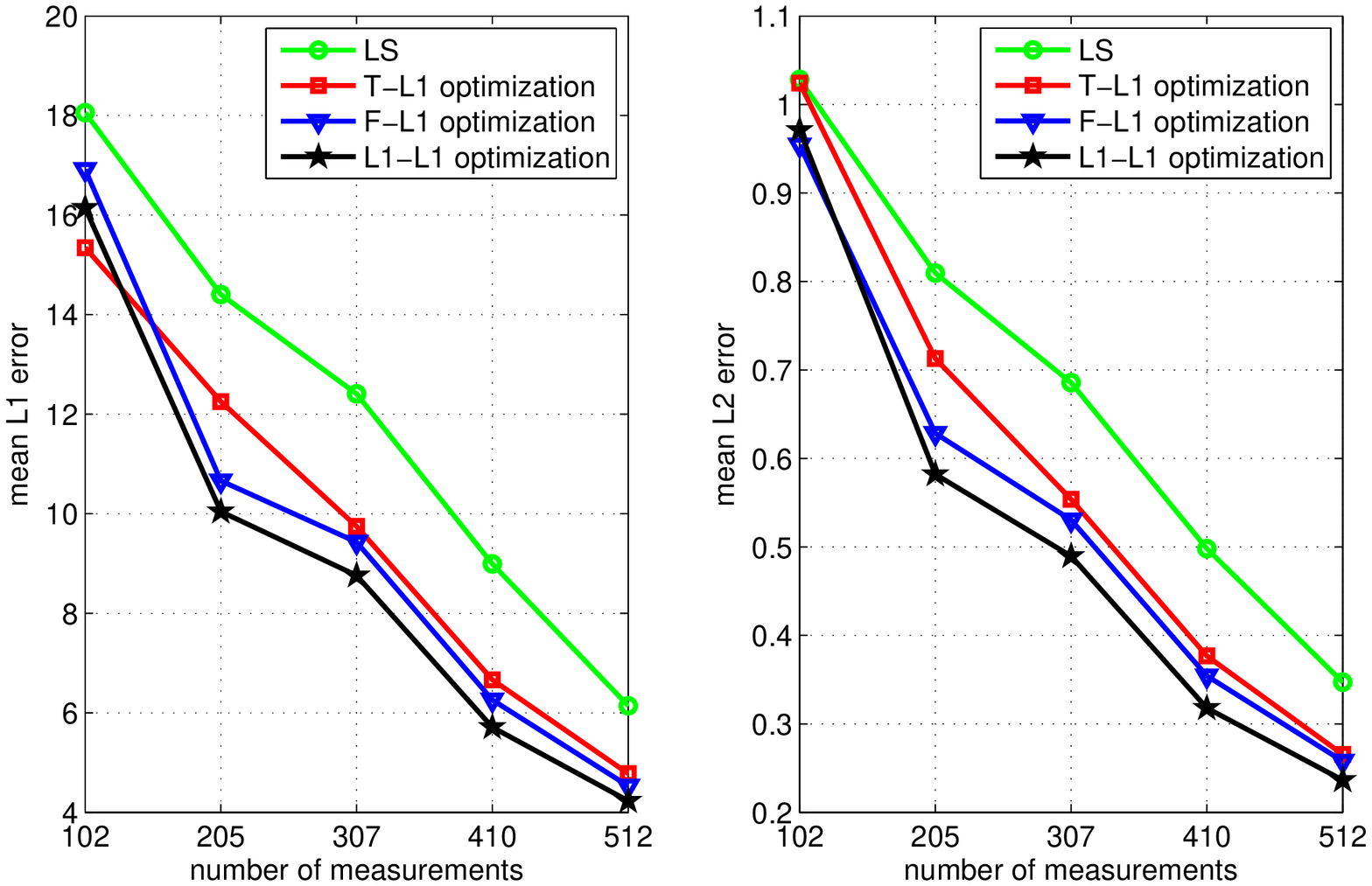}
 \caption{The mean L1 and L2 errors versus the number of measurements with the data $ EMG - healthy $.}
 \label{fig304}
\end{figure}

\begin{figure}[!h]
 \centering
 \includegraphics[scale = 0.32]{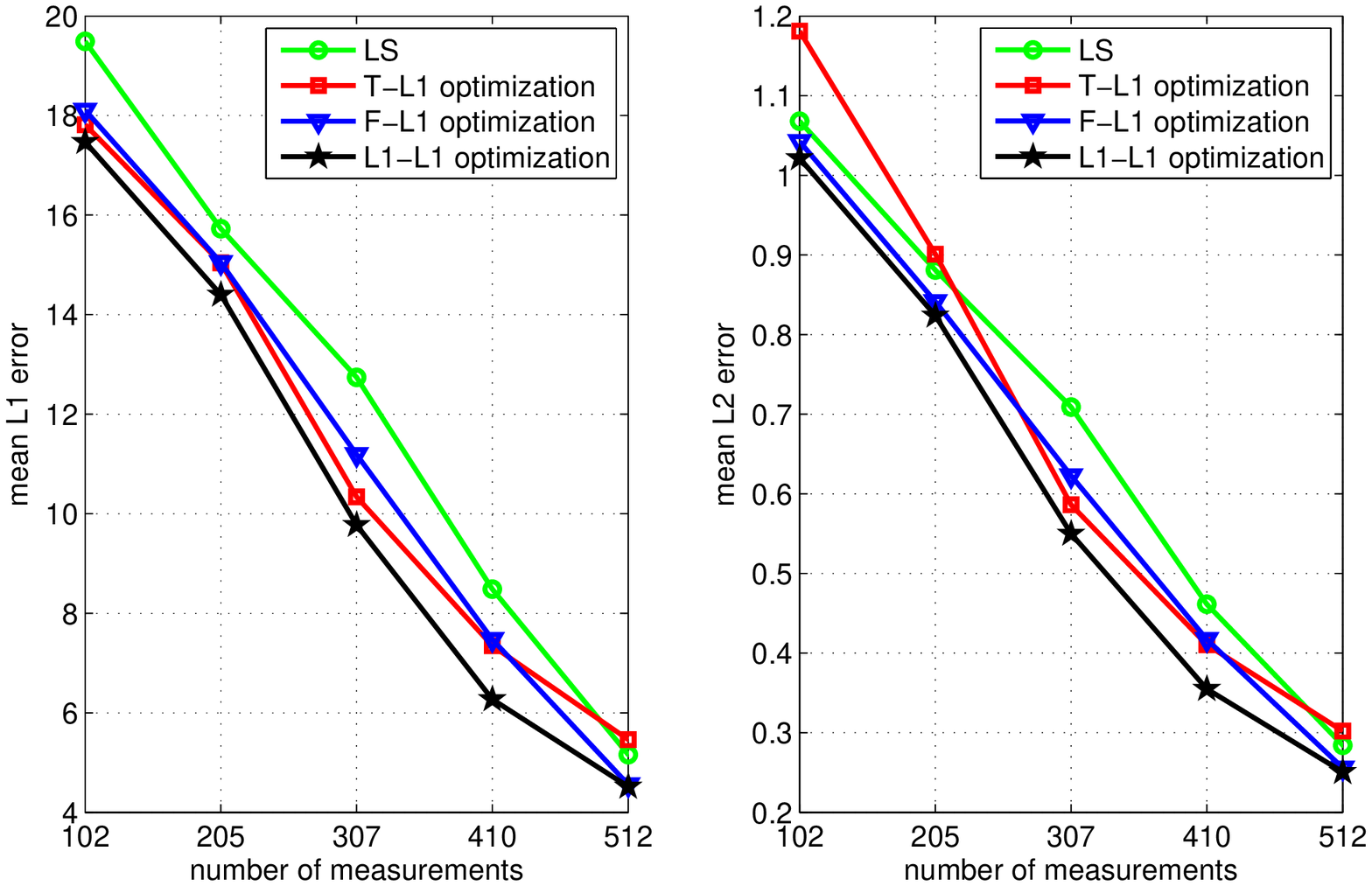}
 \caption{The mean L1 and L2 errors versus the number of measurements with the data $ EMG - myopathy $.}
 \label{fig305}
\end{figure}

\begin{figure}[!h]
 \centering
 \includegraphics[scale = 0.32]{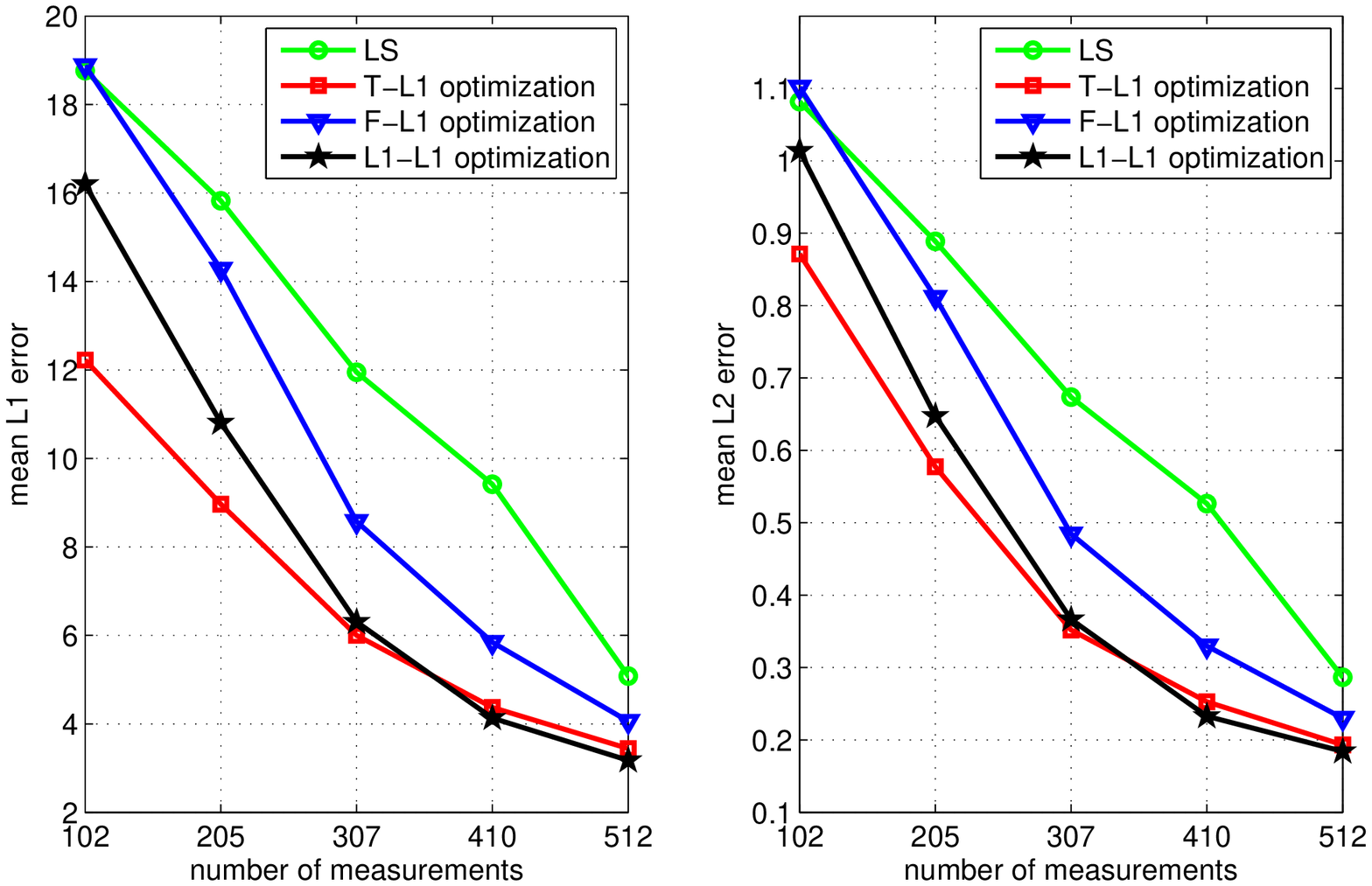}
 \caption{The mean L1 and L2 errors versus the number of measurements with the data $ EMG - neuropathy $.}
 \label{fig306}
\end{figure}

Fig. \ref{fig304}, Fig. \ref{fig305} and Fig. \ref{fig306} show the recovery performance of the three different EMG signals. Here the length of the original EMG signal sections is equal to \emph{N} = 512. All estimation errors decrease with increasing sub-sampling ratio \emph{M/N}. When the sub-sampling ratio reaches 1, the perfect reconstruction with \emph{e} = 0 is still not achieved. This is due to the relaxation of the constraint from $ {\bf{y}} = {\bf{\Phi x}} $ to $ {\left\| {{\bf{y}} - {\bf{\Phi x}}} \right\|_2} \le \varepsilon $. It may be the price for robustness. Besides, because all the EMG data are noisy, and the noiseless signal is not available in (\ref{eq6_vector error}) (\ref{eq6_matrix error}), the performance may be better than demonstrated.

\begin{figure}[!h]
 \centering
 \includegraphics[scale = 0.32]{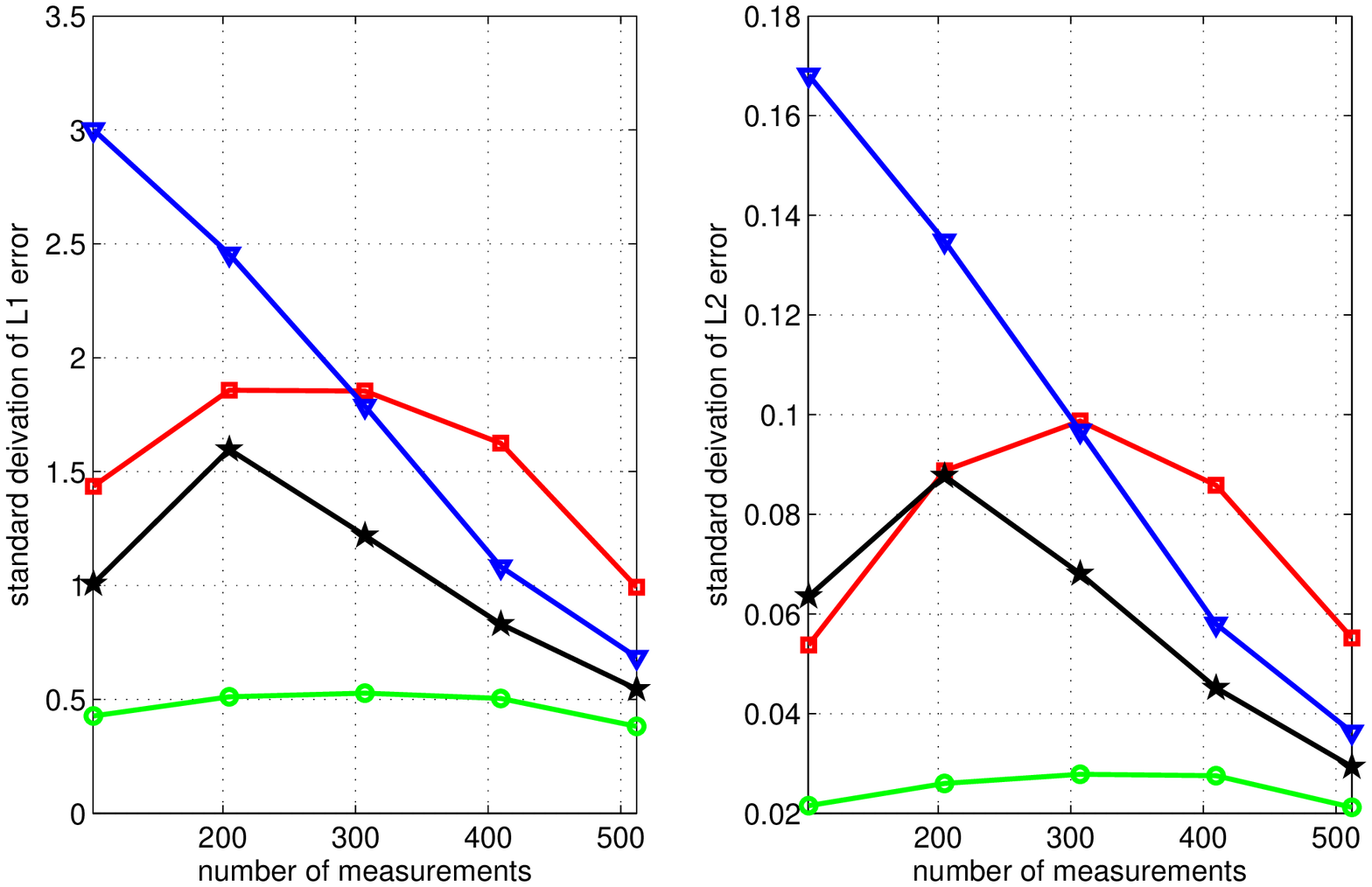}
 \caption{The standard deviation of L1 and L2 errors versus the number of measurements with the data $ EMG - healthy $.}
 \label{fig307}
\end{figure}

\begin{figure}[!h]
 \centering
 \includegraphics[scale = 0.32]{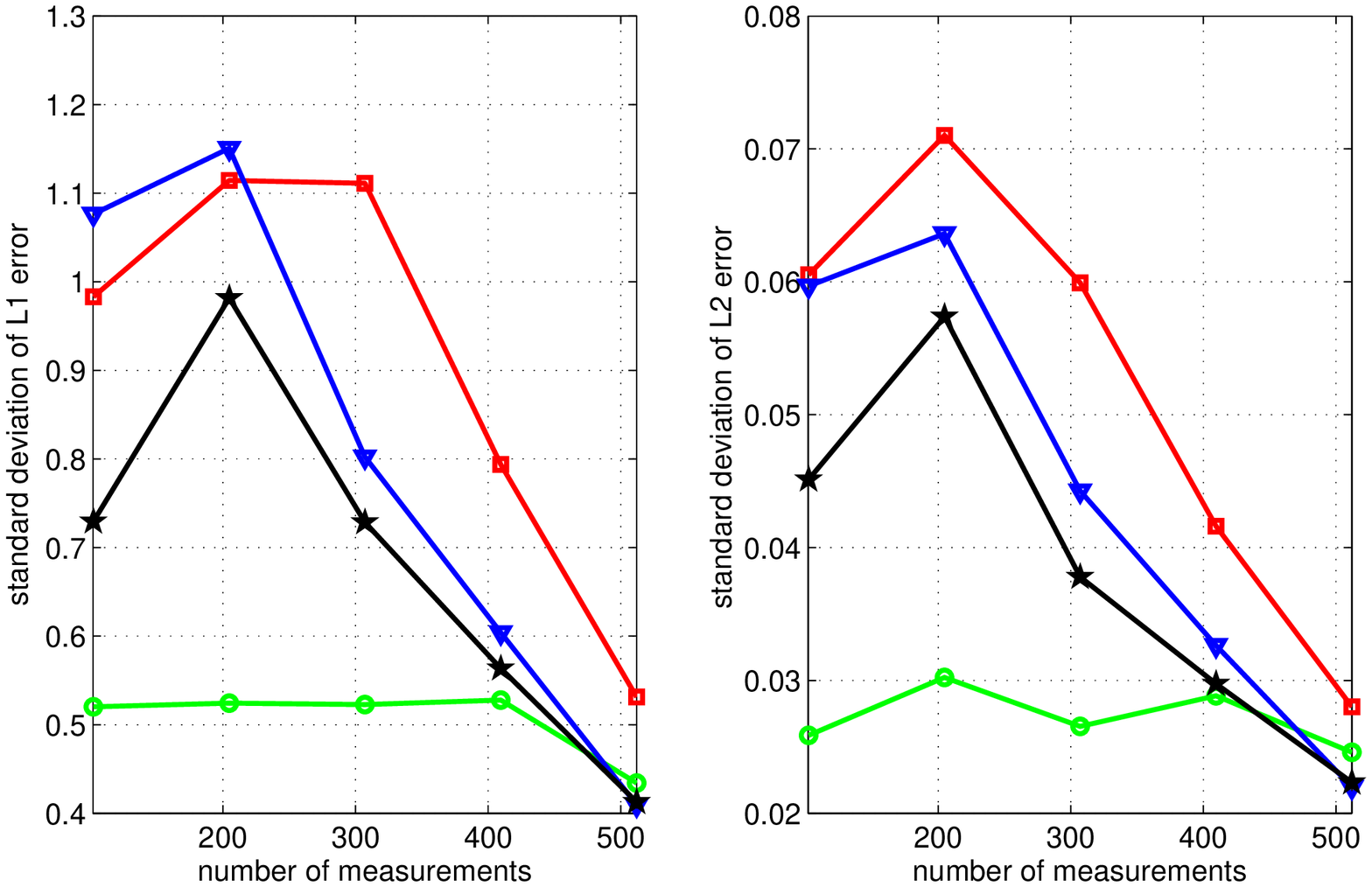}
 \caption{The standard deviation of L1 and L2 errors versus the number of measurements with the data $ EMG - myopathy $.}
 \label{fig308}
\end{figure}

\begin{figure}[!h]
 \centering
 \includegraphics[scale = 0.32]{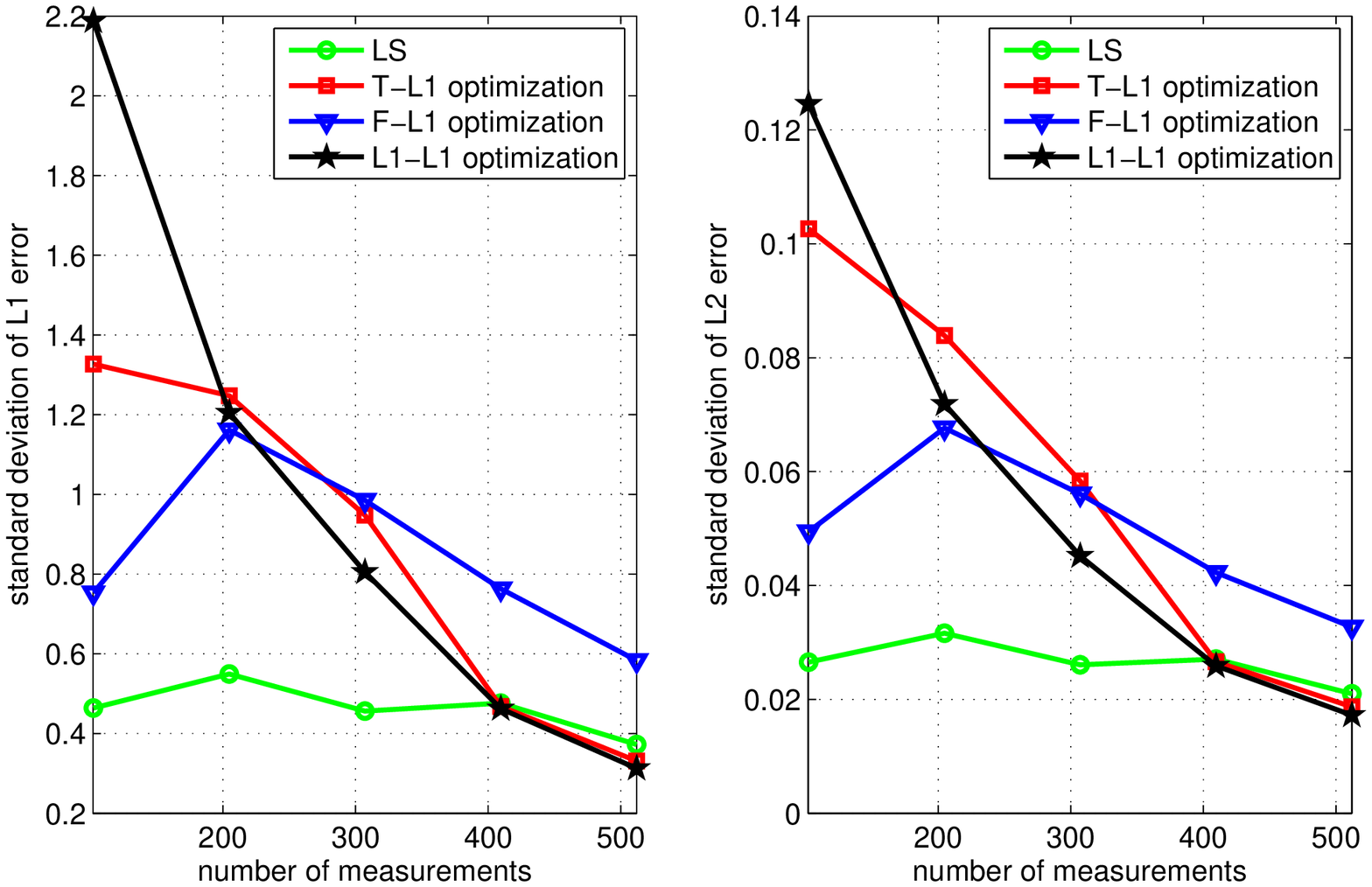}
 \caption{The standard deviation of L1 and L2 errors versus the number of measurements with the data $ EMG - neuropathy $.}
 \label{fig309}
\end{figure}

Fig. \ref{fig307}, Fig. \ref{fig308} and Fig. \ref{fig309} show the standard deviations of EMG signal reconstruction. We can see that the proposed method has a better standard deviation performance than those of the other sparse signal recovery methods in Fig. \ref{fig307}, Fig. \ref{fig308}. In Fig. \ref{fig309}, we can see that the values of the standard deviation of the L1-L1 optimization are smaller than those of the other sparse signal recovery methods except when the number of measurements is smaller than about 200. Despite the fact that the proposed standard deviations of the L1-L1 optimization are larger than those of the LS, the LS method can not be a good candidate for EMG signal recovery from compressive measurements, since the mean L1 and L2 errors of LS are much larger than that of the other three methods.

\begin{figure}[!h]
 \centering
 \includegraphics[scale = 0.38]{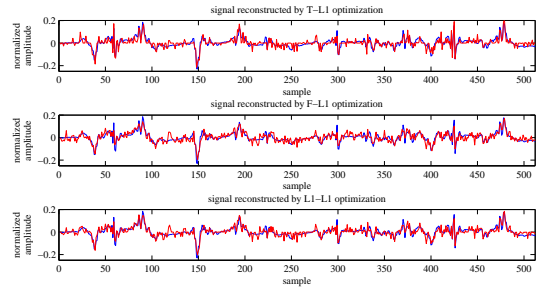}
 \caption{An example of the reconstruction of a section of $  EMG - myopathy $ signal with sub-sampling ratio equals to 0.50. The red ones are the original signals; and the blues ones are the estimated signals. }
 \label{fig3010}
\end{figure}

To illustrate the recovery performance more directly, Fig. \ref{fig3010} shows an example of the reconstruction of a section of $  EMG - myopathy $ signal with sub-sampling ratio equals to 0.50. We can see that the profile of the signal is well reconstructed.

In Fig. \ref{fig304}, T-L1 optimization performs better than F-L1 optimization; but in Fig. \ref{fig305}, F-L1 optimization is better than T-L1 optimization. However, L1-L1 optimization is superior in both Fig. \ref{fig304} and Fig. \ref{fig305}. In Fig. \ref{fig306}, we can see that L1-L1 optimization is better than F-L1 optimization, but worse than T-L1 optimization. The reason is that the EMG signal here is not sparse in the frequency domain, which is evident from Fig. \ref{fig303}.

In summary, if the EMG signal is approximately sparse in both time and frequency domains, L1-L1 optimization is the best candidate for compressive EMG signal recovery. Moreover, if the signal is likely to be sparse in multiple domains with a certain degree of uncertainty, the L1-L1 optimization is also a robust choice, because it can at least avoid the worst performance.

\subsection{MRI}
\label{sec6.4}

\begin{figure}[!h]
 \centering
 \includegraphics[scale = 0.27]{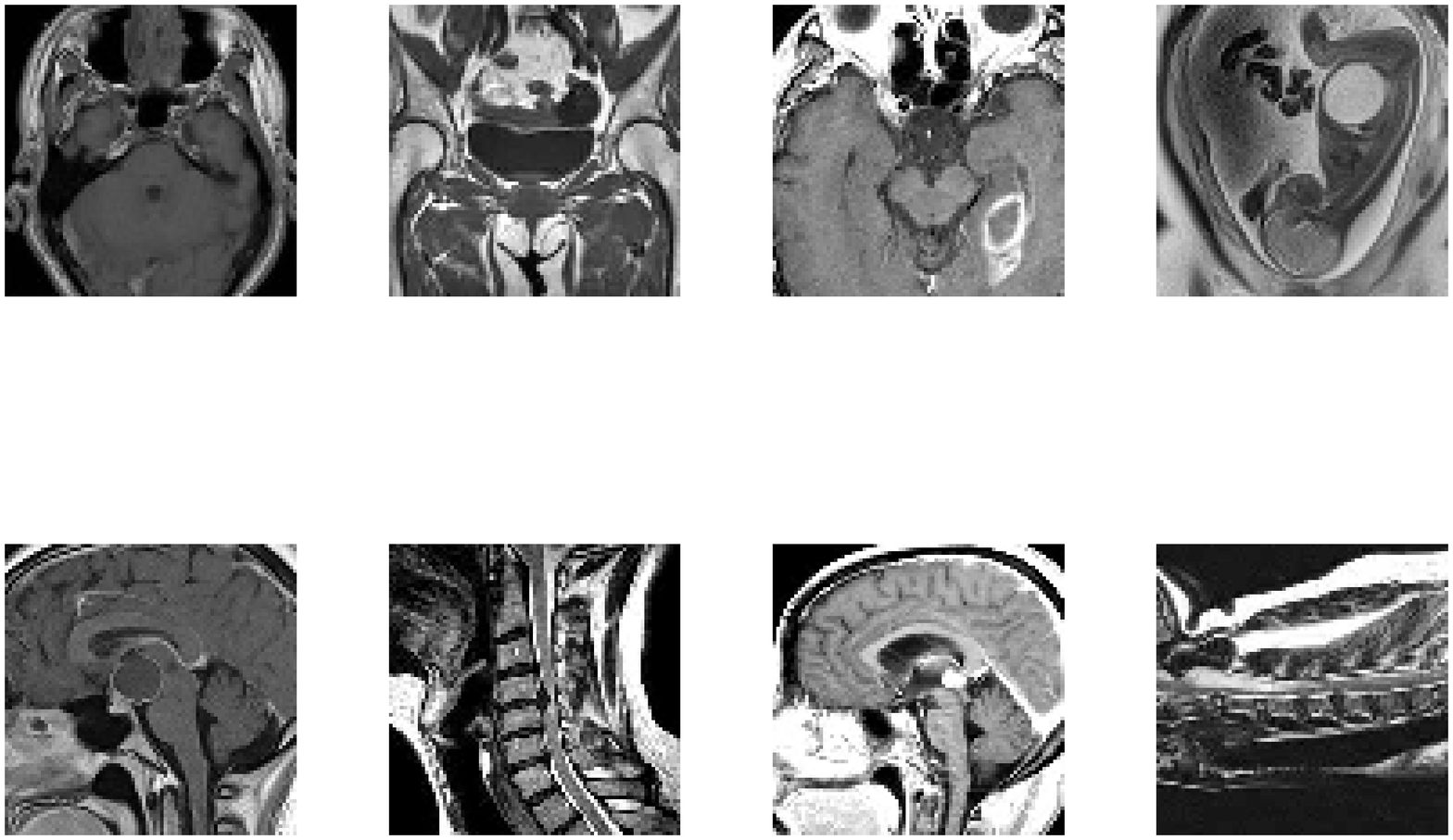}
 \caption{The used medical images \cite{lustig_sparse_mri} \cite{lustig_cs_mri} \cite{romberg_cs}  \cite{needell_tvm}   \cite{lingala_dynamic MRI}.}
 \label{fig401}
\end{figure}

In the experiments, we selected \emph{C} = 8 MRIs with 81-by-81 pixels, as shown in Fig. \ref{fig401}.  The measurement matrix $ {\bf{\Phi }} $  is formed by sampling the i.i.d. entries from a white Gaussian distribution. Because the TV1 is able to recover magnetic resonance images, and the images have low rank structure, BP (\ref{eq3.1 BP}), nuclear norm based recovery (\ref{eq3.3 convex low-rank recovery}), and L1-nuclear optimization (\ref{eq4.3 L1-nuclear optimization}) are used to reconstruct the images. Here the dictionary for sparse representation is \emph{D} with \emph{i}=1 in (\ref{eq3.2 differential matrix}). In L1-nuclear optimization, $ {\lambda _2} $ is chosen to be 3.

%The singular values of the images in Fig. \ref{fig401} are showed in the corresponding position in Fig. \ref{fig402}. We can see that all the singular values of an image are sparsely distributed. The images can be regarded to have low rank.

\begin{figure}[!h]
 \centering
 \includegraphics[scale = 0.37]{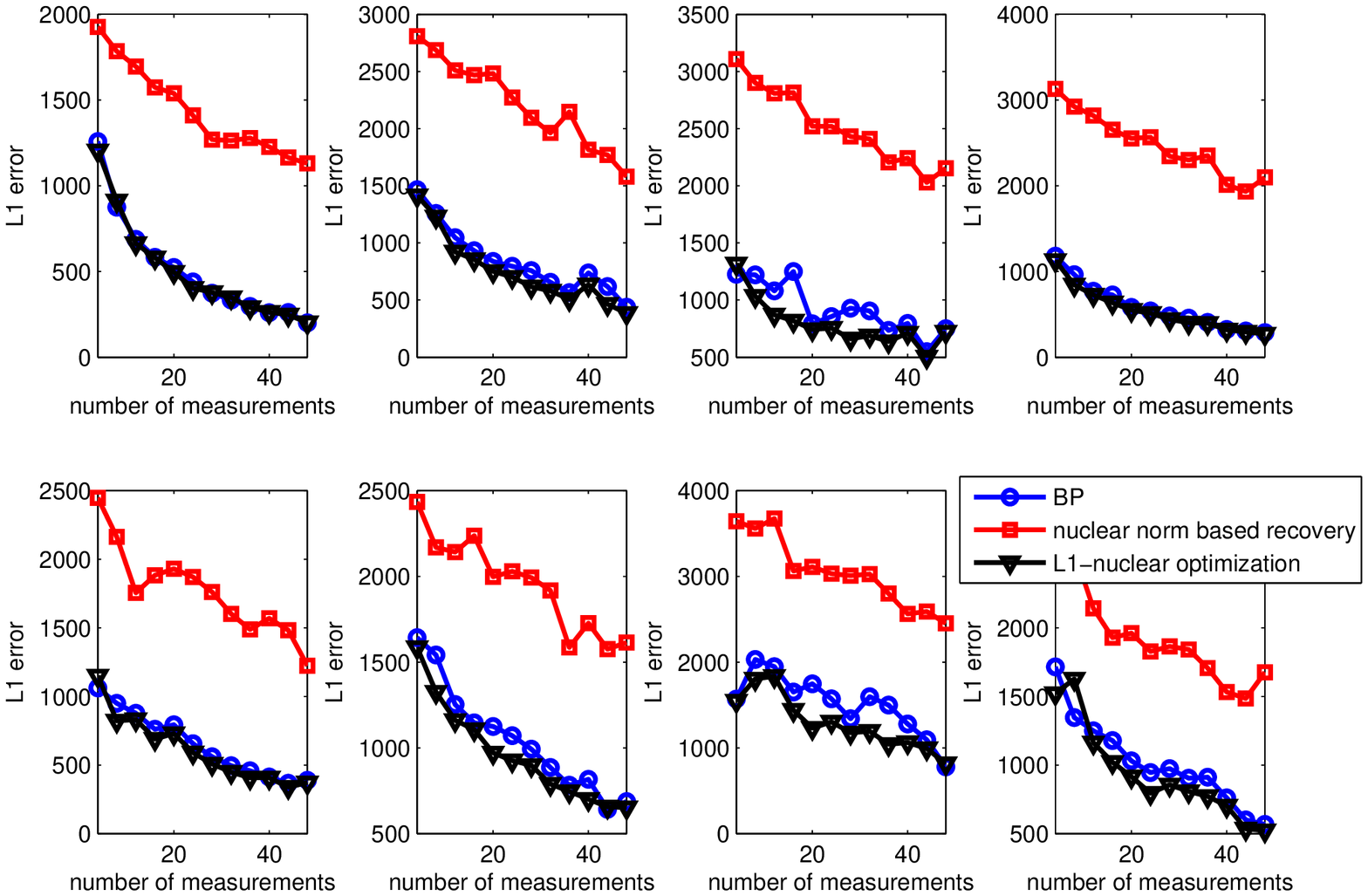}
 \caption{The L1 errors versus the number of measurement when the images in Fig. \ref{fig401} are reconstructed.}
 \label{fig402}
\end{figure}

\begin{figure}[!h]
 \centering
 \includegraphics[scale = 0.37]{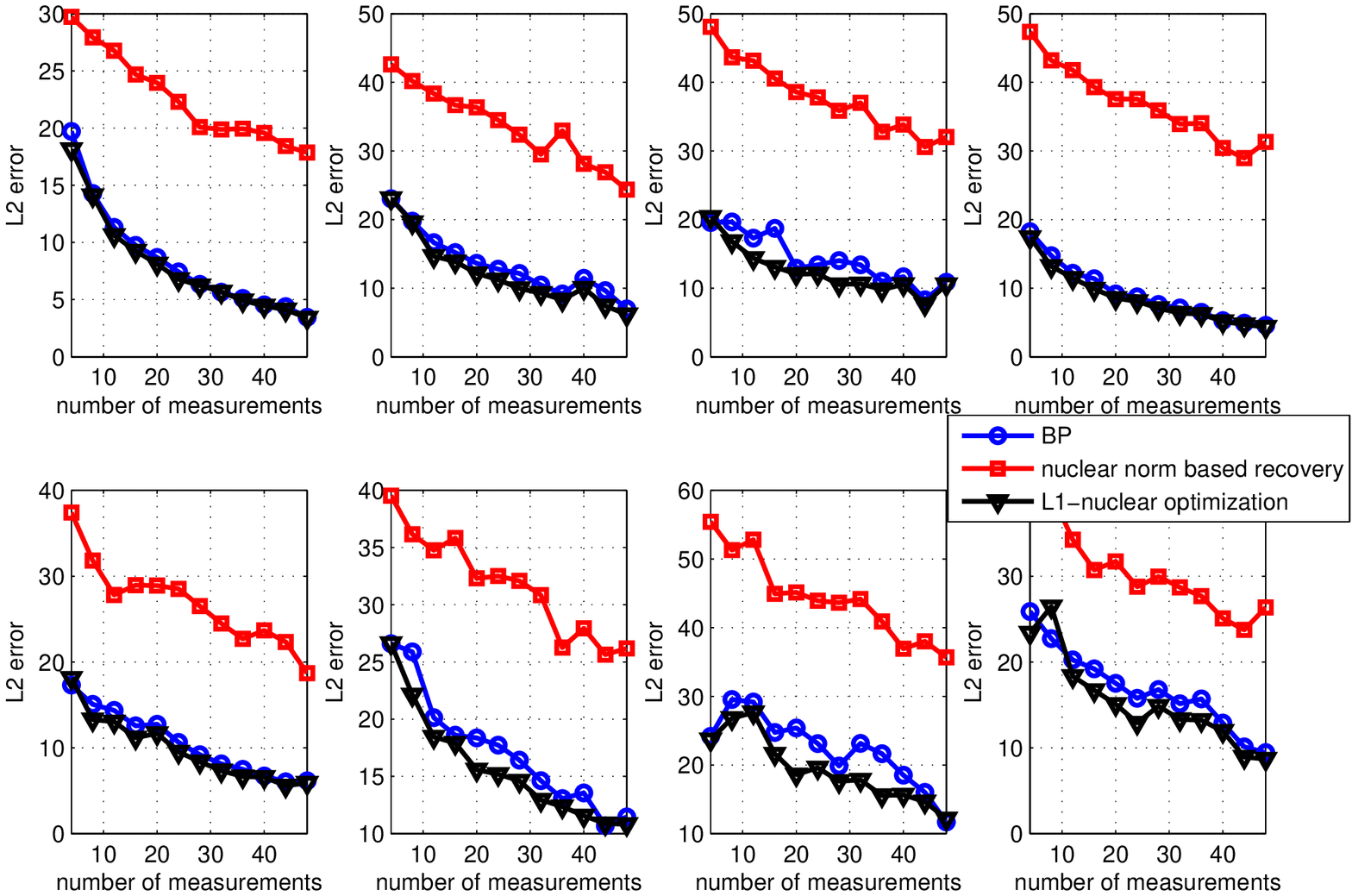}
 \caption{The L2 errors versus the number of measurement when the images in Fig. \ref{fig401} are reconstructed.}
 \label{fig403}
\end{figure}

Fig. \ref{fig402} and Fig. \ref{fig403} show the L1 and L2 errors with different number of sub-sampled measurements when the images in Fig. \ref{fig401} are reconstructed. Every image is normalized by its maximum element. In Fig. \ref{fig402} and Fig. \ref{fig403}, we can see that both the L1 and L2 errors of nuclear norm based recovery are much larger than the ones of BP, which agrees with the fact that nuclear norm minimization constraint is not used for CS, but for matrix completion \cite{candes_matrix_comletion}. In these figures, obviously we can see that the proposed L1-nuclear optimization is better than BP. Although the nuclear norm based recovery has bad accuracy to recover the signal, the nuclear norm minimization for encouraging low rank structure in the estimated matrix can improve the performance of BP which only exploits the sparse structure.

In fact, L1-nuclear optimization has already been used to dynamic MRI. Its performance was shown in \cite{lingala_dynamic MRI}.

\section{Conclusion}
\label{sec7}

In this paper, we give a novel framework for multi-structure signal recovery for CS. The newly proposed methods impose different data structures which are common in biomedical signals. Since more \emph{a priori} information is exploited, the signal recovery performance is enhanced. Numerical experiments confirm the performance improvement.

%
%\section*{Acknowledgment}
%
%This work was supported by Research Council KUL: GOA MaNet, CoEEF/05/006 Optimization in Engineering (OPTEC), PFV/10/002 (OPTEC), IDO 08/013 Autism, several PhD/postdoc and fellow grants; Flemish Government: FWO: PhD/postdoc grants, projects: FWO G.0302.07 (SVM), G.0341.07 (Data fusion), G.0427.10N (Integrated EEG-fMRI), G.0108.11 (Compressed Sensing) G.0869.12N (Tumor imaging) research communities (ICCoS, ANMMM); IWT: TBM070713-Accelero, TBM070706-IOTA3, TBM080658-MRI (EEG-fMRI), PhD Grants; IBBT; Belgian Federal Science Policy Office: IUAP P6/04 (DYSCO, 'Dynamical systems, control and optimization', 2007-2011); ESA AO-PGPF-01, PRODEX (Cardio Control) C4000103224 EU: RECAP 209G within INTERREG IVB NWE programme, EU HIP Trial FP7-HEALTH/ 2007-2013 (n 260777) ( Neuromath (COST-BM0601); BIR$ \& \ $D Smart Care; Alexander von Humboldt stipend.
%

\ifCLASSOPTIONcaptionsoff
  \newpage
\fi

\clearpage

% that's all folks
\end{document}